\documentclass[prb,twocolumn,superscriptaddress,longbibliography,aps]{revtex4-1}

\usepackage[T1]{fontenc}
\usepackage[utf8]{inputenc}
\usepackage{lmodern}
\usepackage[sumlimits]{amsmath}
\usepackage{amssymb}
\usepackage{bm}
\usepackage{yhmath}
\usepackage{graphicx}
\usepackage[no-test-for-array]{nicematrix}
\usepackage{epstopdf}
\usepackage{latexsym}
\usepackage{color}
\usepackage{nicefrac}
\usepackage{dsfont}
\usepackage{bbold}
\usepackage{wasysym} 
\usepackage{stmaryrd}
\usepackage{hyperref} 
\usepackage{xcolor}
\usepackage{multirow}
\usepackage{soul}
\usepackage{adjustbox}
\usepackage{cancel}

\usepackage{esint}
\usepackage{tikz}
\usepackage{mathtools}

\newcommand{\be}{\begin{equation}}
\newcommand{\ee}{\end{equation}}
\newcommand{\bea}{\begin{eqnarray}}
\newcommand{\eea}{\end{eqnarray}}

\newcommand{\idmatrix}{\text{\textbb{1}}}

\usepackage{verbatim}
\usepackage{dsfont} 
\usepackage{caption}
\usepackage{url}

\usepackage{subcaption}

\usepackage{sansmath}
\DeclareFontEncoding{LGR}{}{}
\DeclareSymbolFont{sfgreek}{LGR}{cmss}{m}{n}
\SetSymbolFont{sfgreek}{bold}{LGR}{cmss}{bx}{n}
\DeclareMathSymbol{\sxi}{\mathord}{sfgreek}{`x}
\DeclareMathSymbol{\stheta}{\mathord}{sfgreek}{`j}
\DeclareMathSymbol{\sepsilon}{\mathord}{sfgreek}{`e}
\DeclareMathSymbol{\sOmega}{\mathalpha}{sfgreek}{`W}
\DeclareMathSymbol{\stau}{\mathalpha}{sfgreek}{`t}

\newcommand{\bs}{\boldsymbol}

\newcommand{\mc}{\mathcal}

\usepackage{scalerel}

\usepackage{accsupp}

\begin{document}

\title{Dynamical magnetic breakdown and quantum oscillations from hot-spot scattering}

\author{L\'eo Mangeolle}
\affiliation{Technical University of Munich, TUM School of Natural Sciences, Physics Department, 85748 Garching, Germany}
\affiliation{Munich Center for Quantum Science and Technology (MCQST), Schellingstr. 4, 80799 M{\"u}nchen, Germany}
\author{Johannes Knolle}
\affiliation{Technical University of Munich, TUM School of Natural Sciences, Physics Department, 85748 Garching, Germany}
\affiliation{Munich Center for Quantum Science and Technology (MCQST), Schellingstr. 4, 80799 M{\"u}nchen, Germany}

\date{\today}
\begin{abstract}
Quantum oscillations (QO) are a well-established probe of Fermi-surface (FS) geometry and in the presence of long-range density wave order can display new QO frequencies from reconstructed FS pockets. We show that such reconstructed  frequencies can arise even in the absence of long-range density order. Considering electrons coupled to a fluctuating bosonic mode that scatters quasiparticles between sharp hot spots on the FS, we develop a semiclassical theory in which the interaction generates time-dependent tunneling processes analogous to magnetic breakdown. This dynamical magnetic breakdown produces new semiclassical orbits corresponding to reconstructed FS areas despite the absence of static order. Because tunneling probabilities depend on the thermal population of bosonic excitations, the resulting oscillation amplitudes exhibit characteristic deviations from standard Lifshitz–Kosevich behavior. Our results  provide a mechanism to probe bosonic fluctuations in quantum critical metals and provide a framework for dynamical magnetic breakdown.
\end{abstract}

\maketitle

\section{Introduction}
\label{sec:introduction}

\begin{figure}[]
  \begin{center}
    \includegraphics[width=\columnwidth]{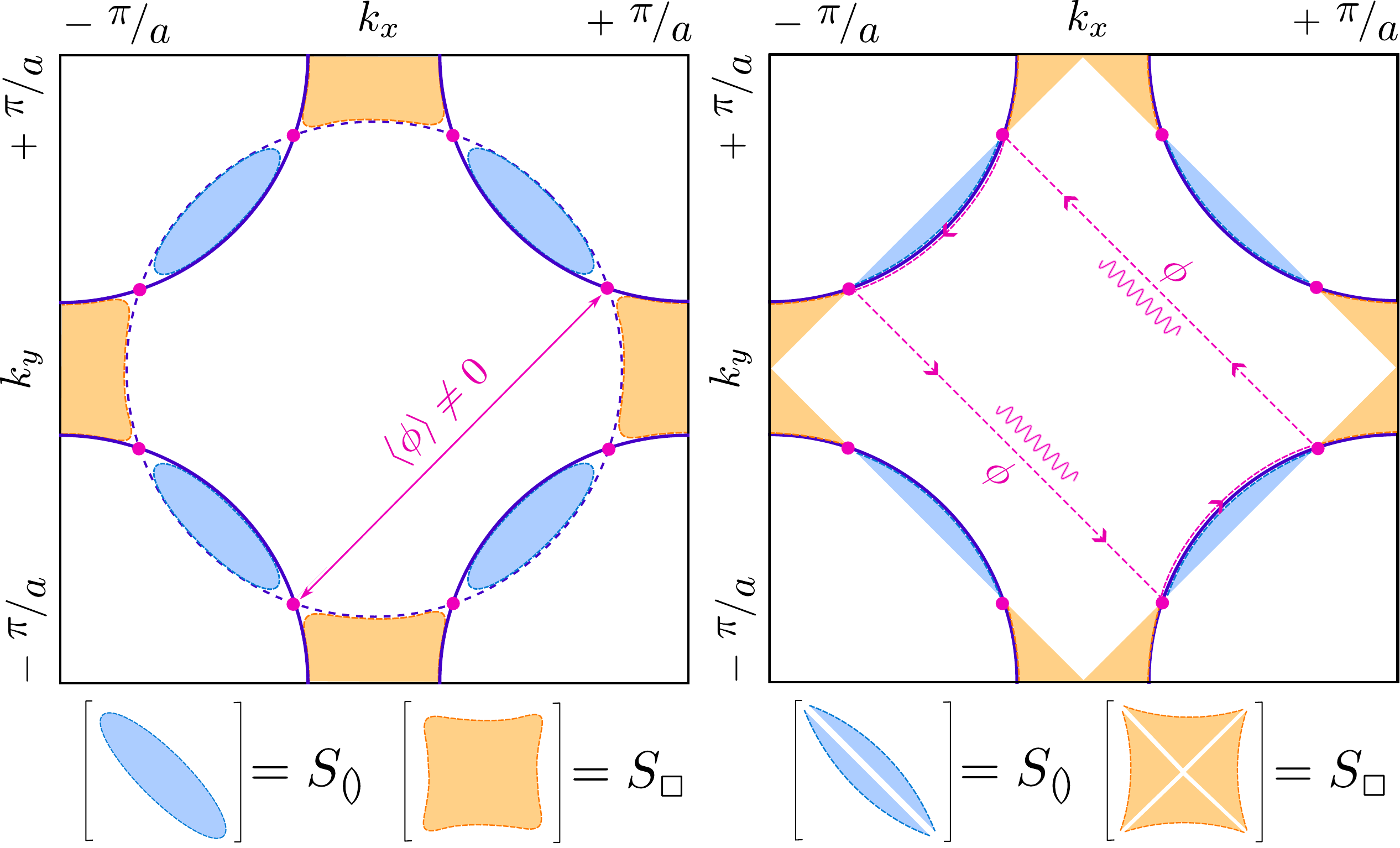}
    \caption{Left: Reconstruction scenario, where the large (deep purple) hole FS is folded in the presence of long-range density wave order and $\langle \phi \rangle \neq 0$ opens gaps at the (pink) hot spots,
      with hole-like (light blue) and electron-like (orange) pockets. The two resulting FS areas proportional to QO frequencies are sketched below.  Right: Dynamical scenario without long-range order, where a fluctuating boson $\phi$ (with $\langle \phi \rangle=0$)
    scatters holes between hot spot pairs, allowing new semiclassical trajectories (e.g. the dashed pink orbit). The effective areas enclosed by the breakdown paths are sketched below.}
\label{fig:advertisement}
\end{center}
\end{figure}

Magneto-oscillations or ``quantum'' oscillations (QO) have long been a standard way of probing Fermi surfaces (FS) of metals \cite{pippard1960experimental,shoenberg1984magnetic}. Their success relies on the fact that their frequency, as a function of $1/B$, is proportional to the FS cross-section transverse to the magnetic field, which allows for FS tomography or `Fermiology'~\cite{alexandradinata2023fermiology}. In turn, it becomes possible to witness important changes in the FS geometry, such as Lifshitz transitions \cite{lifshitz1960anomalies} and FS reconstruction induced by the development of long-range density wave order quantified by an order parameter \cite{PhysRev.128.1437,hertz}. Such reconstruction, attributed in particular to a (spin) density wave instability \cite{monthoux1991toward,PhysRevB.51.14874,pines1996nearly}, was observed experimentally for example in the QO spectrum of the parent compound cuprate superconductors \cite{doiron2007quantum, doi:10.1073/pnas.0804002105,Sebastian_2012,sebastian2015quantum}, in that of iron-based superconductors \cite{Sebastian_2008, Carrington_2011}, or heavy fermion systems~\cite{hassinger2010similarity,shishido2009possible}.
The basic 2D picture is that of Fig.\ref{fig:advertisement} (left), where the folding of a large FS with area $S_{\text \Circle}$, upon development of an order parameter $\langle \phi \rangle \neq 0$, reconstructs it into a new set of Fermi pockets with different areas $S_{(\!)}$ and $S_{\text \Square}$. Crucially, it implies the appearance of new QO frequencies (and the disappearance of that associated with $S_{\text \Circle}$). Numerous theory treatments of QOs in that context exist, see e.g. \cite{allais2014connecting} and references therein. Normally, the formation of an order parameter $\langle \phi \rangle \neq 0$ is assumed as a prerequisite for the apperance of new reconstructed QO frequencies, and the effect of bosonic fluctuations of $\phi$, investigated \textit{a posteriori} within a perturbation theory framework, merely renormalizes the reconstructed QO spectrum and amplitudes. 

Here, we address a key unsolved question, namely, whether the appearance of a new QO frequency coincides with the phase transition to long-range density wave order, or precedes it. In addition, we discuss how critical fluctuations influence the QO amplitudes prior to the transition. 
The first question belongs to the more general problem of how interactions and scattering can generate new QO frequencies that are not simply related to free FS orbits, generally dubbed as non-Onsager frequencies \cite{leeb2025field}. On the one hand, it has been established that unexpected QOs can appear in inverted insulators without FSs whatsoever~\cite{knolle2015quantum,knolle2017anomalous,allocca2022quantum}. On the other hand, unconventional QO can also appear in multiband metals as a result of interband scattering mediated by a fluctuating boson \cite{mangeolle2025anomalous}, which in particular display unusual temperature dependence. However, existing theories always found new frequencies as combinations of the existing frequencies\footnote{The fine structure of frequency peaks can also involve the boson's frequency, cf Ref.\cite{mangeolle2025anomalous}.} (in particular the highly unusual difference frequency \cite{leeb2023theory}, or the related experimental obervation of QO of the quasiparticle lifetime~\cite{huber2023quantum}), unlike those generated in a reconstruction scenario. A theory for how the latter can emerge from near-critical bosonic fluctuations has been missing.

The second question must be considered in the context of long-standing results on properties of a Fermi liquid coupled to bosonic excitations. In particular, the influence of electron-phonon coupling on the amplitude of QOs  has been studied long time ago~\cite{fowler1965electron,engelsberg1970influence} constraining the allowed temperature dependences. We address, whether these ``no-go theorems'' are obeyed in the pre-reconstruction setting we will consider here.

A crucial ingredient in our study will be the existence of singular electron-boson scattering at sharp isolated FS 'hot spots'. This allows to sharply define new semiclassical orbits, see Fig.\ref{fig:advertisement} (right), similar to those appearing in the reconstruction scenario. The central result of our work is to show that these semiclassical orbits exist, even without assuming boson condensation (i.e.  formation of long-range order).  They appear directly in the QO spectrum. Moreover, we identify an intermediate magnetic breakdown regime where both sets of oscillations (original, and reconstructed) exist simultaneously, with distinct temperature dependence beyond the standard Onsager-Lifshitz-Kosevich paradigm \cite{onsager1952interpretation, lifshitz1956theory}.

The rest of the paper is organized as follows. In section \ref{sec:setup}, we introduce our (standard) electron-boson model with hot spots, and define notations. In sections \ref{sec:saddle-point-appr}-\ref{sec:geometry-hot-spot}, we perform a saddle-point approximation to derive an effective problem of electrons coupled to a time-dependent background. In sections \ref{sec:effect-tunn-probl}-\ref{sec:pert-analyt-solut}, we solve this problem in a magnetic field semiclassically, as a time-dependent tunneling problem. In section \ref{sec:orbital-quantiz-phys-conseq}, we compute physical quantities from this semiclassical picture, and discuss experimental implications.
Throughout the paper we use units where $\hbar = 1 = k_{\rm B}$.

\section{Setup}
\label{sec:setup}

We consider a single large Fermi surface of electrons (or holes), coupled to a dynamical bosonic field.
The free-particle Hamiltonians is
\begin{align}
  \label{eq:27}
  H_f = \int_{\bs k} \epsilon_{\bs k} c^\dagger_{\bs k} c_{\bs k} , \quad  H_b = \int_{\bs q} \omega_{\bs q} b^\dagger_{\bs q} b_{\bs q}, 
\end{align}
and their coupling reads, in full generality,
\begin{align}
  \label{eq:28}
 H^{(0)}_{\rm int} &= \int_{\bs k, \bs q} g_{\bs q}^{(0)} c^\dagger_{\bs k+\bs q} c_{\bs k} b_{\bs q} + \rm h.c. ,
\end{align}
where $g_{\bs q}^{(0)}$ is the bare coupling constant, which depends on the nature of the collective mode
(e.g. the boson could be a spin-density or charge-density fluctuation, a soft phonon, etc) and details of its coupling to electrons.

We now assume that the boson dispersion has a global minimum $\omega_{\bs Q}$ at momentum $\bs Q=(\pi/a,\pi/a)$ connecting the 'hot spots' on the Fermi surface, with $a$ the square lattice parameter.
This  scenario is well established to originate from (quasi-)nesting situations and increased scattering with selected momenta~\cite{Sachdev_2011}, as represented in Fig.\ref{fig:advertisement}.

The effective low-energy theory of the coupling \cite{abanov2003quantum} involves only electrons located near each hot spot $\bs K$,
described as fermionic fields $\psi_{\bs K}(\bs k)$ (whose conjugate in functional integral notations is $\bar \psi_{\bs K}(\bs k)$)
coupled to a dynamical bosonic field $\phi$ (with conjugate $\bar \phi$). 
To be more precise, defining $\psi_{\bs K}(\bs k) = \left ( c_{\bs{K+k}} , c_{\bs{K+k+Q}} \right )^\top$,
the fermion sector of the theory  may always be rewritten as
\begin{align}
  \label{eq:29}
  \mc L_f &= \frac 1 2 \sum_{\bs K \in {\rm HS}} \int'_{\bs k}  \bar \psi_{\bs K}(\bs k) \, i\partial_t \psi_{\bs K}(\bs k) - H_f, \\
  H_f &= \frac 1 2 \sum_{\bs K \in {\rm HS}} \int'_{\bs k} \bar \psi_{\bs K}(\bs k)
         \begin{bmatrix}
    \epsilon_{\bs{K+k}} & 0 \\ 0 &  \epsilon_{\bs{K+k+Q}} 
  \end{bmatrix}
\psi_{\bs K}(\bs k) ,
\end{align}
where HS is the set of hot spots and $\int_{\bs k}'$ (relative to a given $\bs K$) integrates over all momenta $\bs K+\bs k$ whose nearest hot spot is $\bs K$.
In a simplified model, the low-energy theory involves one single bosonic field, with Hamiltonian $H_\phi = \Omega \,\bar \phi \phi$ (with the boson frequency $\Omega$),
and the interaction
\begin{align}
  \label{eq:30}
  H_{\rm int} &= \frac 1 2 \sum_{\bs K \in {\rm HS}} \int'_{\bs k} \left (  \phi  \, g(\bs k) + \bar \phi  \,  g(\bs k)^*\right )\;
                \bar \psi_{\bs K}(\bs k)
     \, \hat \sigma^x\, \psi_{\bs K}(\bs k) ,
\end{align}
where $\hat \sigma^x$ is the first Pauli matrix.
Here, $ g(\bs k) $ is an \emph{effective} coupling constant
that takes finite values around $\bs k = \bs 0$ and decreases at larger momenta, that is to say, going away from the hot spot.
The derivation of this effective theory is described with more detail in Appendix \ref{sec:pert-deriv-effect}, as it is not our focus here.
The precise form of $ g(\bs k) $, as derived from $g_{\bs {K+k}}^{(0)}$, can in principle be determined perturbatively from vertex renormalization,
but in the following we will simply assume a convenient form, see Eq.\eqref{eq:73}.
We now use Eq.\eqref{eq:30} as a starting point, and investigate the semiclassical dynamics of electrons in the presence of such an interaction.

\section{Semiclassical solution}
\label{sec:semicl-solut}

\subsection{Saddle-point approximation}
\label{sec:saddle-point-appr}

We study semiclassical quantum oscillations, which involve the physics of electrons at the Fermi surface.
Their interaction with a boson $\phi$, that may describes collective electronic modes or phonons in an effective way, is therefore \emph{elastic} at the semiclassical level. 
In that case, it is possible to treat the boson at the saddle-point (i.e. as a classical field) and evaluate properties of Fermi surface electrons in this classical background,
eventually averaging over bosonic configurations. This approach has been successfully applied to the physics of weak localization \cite{BLAltshuler_1982,Aleiner01041999},
of SDW fluctuations \cite{kampf1990pseudogaps} and of gauge field fluctuations \cite{khveshchenko1993low} in doped antiferromagnets,
and of phase fluctuations in doped $d$-wave superconductors \cite{franz1998phase}.

Thus, within the full Lagrangian of the problem,
\begin{align}
  \label{eq:74}
  \mc L = \mc L_f - H_{\rm int} +  \bar \phi (i\partial_t - \Omega) \phi ,
\end{align}
we perform a saddle-point approximation of the bosonic fields, replacing $\phi(t) \rightarrow \phi_{0}\,e^{-i\Omega t}$
and $\bar \phi(t) \rightarrow \phi_{0}^*\,e^{+i\Omega t}$ in the fermion's functional integral,
where $\phi_{0}$ is a complex \emph{number} that still needs to be averaged over.
The precise formulation of this saddle-point approximation within the full functional integral description of the problem is discussed in Appendix \ref{sec:saddle-point-appr-1}.
Physically, this assumption requires that the back-action of electrons on the boson dynamics is negligible,
which can ensue, for instance, from Migdal's theorem \cite{migdal1958interaction} or a large-$N$ approximation
(notably the number of hot spots in the geometry of Fig.\ref{fig:advertisement} can be used as $N$, see Ref.\cite{abanov2003quantum}).

Crucially, this is not equivalent to condensing the bosonic field,
because fluctuations are still accounted for by the average over $\phi_{0}, \phi_{0}^*$.
Rather, in the present case, bosonic fluctuations are only assumed to retain coherence over time scales much larger
than the typical electronic scales appearing in dynamical properties, e.g.\ the cyclotron period $2\pi/\omega_{\rm c}$,
but the bosonic field does not have to be static or even slow (in particular, $\Omega/\omega_{\rm c}$ need not be small).
Thus, contrary to the Fermi surface reconstruction scenario where the boson indeed condenses and opens gaps at the hot spots (see Fig.\ref{fig:advertisement}),
here we investigate the case where the boson is dynamical and fluctuates, which we treat in the saddle-point approximation.

The resulting theory is a collection of independent problems, indexed by a complex number $\phi_{0}$,
such that for a given fixed $\phi_{0}$ the theory has become purely electronic, and is that of free electrons with a time-dependent Lagrangian:
\begin{align}
  \label{eq:93}
   \mc L_{\rm eff} &= \frac 1 2 \sum_{\bs K \in {\rm HS}} \int'_{\bs k} \bar \psi_{\bs K}(\bs k) \,\left (   i\partial_t \idmatrix - H_{\bs{K}}(\bs k) \right )\,
                                                                                           \psi_{\bs K}( \bs k)  , 
\end{align}
where $\idmatrix$ is the two-by-two identity matrix, and
\begin{align}
  \label{eq:75}
  H_{\bs{K}}(\bs k,t) =
  \begin{bmatrix}
\epsilon_{\bs {K+k}} & \Delta (\bs k,t) \\ \Delta (\bs k,t)  &   \epsilon_{\bs{K+k+Q}} 
\end{bmatrix}    \equiv H_{\rm eff}   ,                                                
\end{align}
where
\begin{align}
  \label{eq:107}
  \Delta (\bs k,t) &=  g(\bs k)^* \, \phi_0^* \,e^{+i\Omega t} +  g(\bs k) \,\phi_0 \,e^{-i\Omega t} 
\end{align}
is an effective time-dependent potential felt by the electrons.
We stress that the latter depends on the bosonic field's configuration at the saddle-point,
and ultimately will always be averaged over to extract physical quantities.

Our objective is to study semiclassical dynamics of electrons in such a time-dependent Hamiltonian, Eq.\eqref{eq:93} in a fixed $\phi_0$ sector.
The semiclassical dynamics of electrons is governed by their Fermi surface properties.
Here we will be focusing on a subset of observables (namely, quantum oscillations
within the standard Onsager \cite{onsager1952interpretation}, Lifshitz-Kosevich \cite{lifshitz1956theory} picture)
that are dictated by the Fermi surface \emph{geometry} (as regards the oscillations' frequency) and \emph{average mass} (as regards their temperature dependence).

\subsection{Geometry of hot spot pairs}
\label{sec:geometry-hot-spot}

We note that because $\Delta (\bs k,t) $ is only significant close to the hot spots, its primary effect will be to generate time-dependent dynamics connecting the hot spots pairwise.
In particular, in the FS reconstruction picture, $\Delta (\bs k,t) $ opens a time-dependent gap at the hot spots,
which strongly affects the (instantaneous) Fermi surface geometry -- and more generally (away from that picture) the instantaneous semiclassical orbits' geometry.
Meanwhile, insofar as the hot spot regions are small compared with the total Fermi surface perimeter, the average mass is little affected.

For future use, we note that it is easy to see (see Appendix \ref{sec:pedestrian} for details) that only the squared modulus 
\begin{align}
  \label{eq:483}
  \left | \Delta (\bs k,t)  \right |^2 &= \tfrac 1 {2} \left | g(\bs k)\right |^2\, \phi_0^* \phi_0 \,\left [  1  + \cos\left ( 2\Omega t - 2 \alpha_0 \right ) \right ] ,
\end{align}
where $ \alpha_0 = {\rm arg} \left [ g(\bs k) \,\phi_0 \right ]$, appears in the effective dynamics at the hot spots.

\begin{figure}[htbp]
  \begin{center}
       \includegraphics[width=\columnwidth]{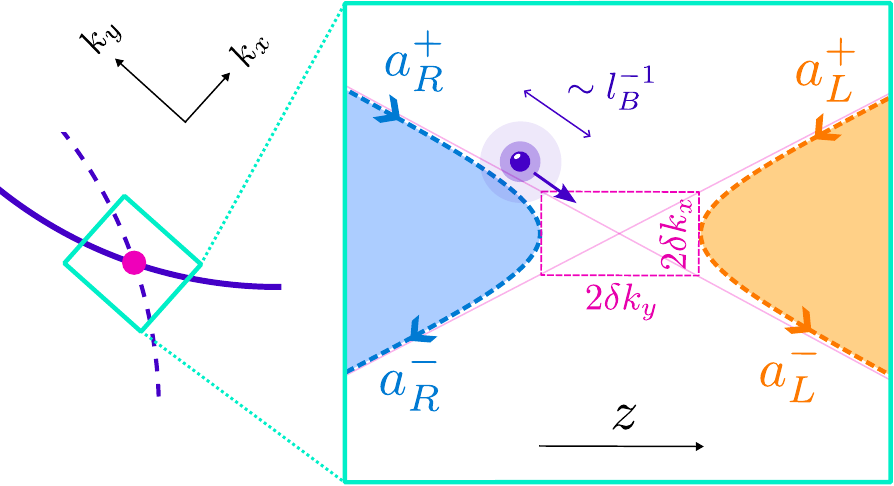}
       \caption{Tunneling geometry at a hot spot. In a given $(\phi_0^*,\phi_0)$ sector and at fixed $t$, the local FS geometry is hyperbolic
         with parameters $(\delta k_x,\delta k_y)$. The local system of axes $(k_x,k_y)$ and tunneling coordinate $z$ are those of Eqs.\eqref{eq:99}.
      The tunneling of a (deep purple) particle through the junction can be described by a relation like Eq.\eqref{eq:115} between wavefunction amplitudes $a_{L/R}^\pm$.}
\label{fig:gap_geometry}
\end{center}
\end{figure}

We now focus on one given pair of hot spots, whose geometry in the FS reconstruction picture is depicted in Fig.\ref{fig:gap_geometry}.
The local geometry is that of two hyperbolae, that can be parameterized in terms of
\begin{subequations}
  \label{eqs:1089}
\begin{align}
  \label{eq:1089a}
  \delta \! k_x &=k_{\rm F}^2/{\rm q} \times \Delta/E_{\rm F} ,\\
    \label{eq:1089b}
   \delta \! k_y &= k_{\rm F}^2 /   \sqrt{k_{\rm F}^2 -{\rm q}^2}  \times \Delta/E_{\rm F}  ,
\end{align}
\end{subequations}
where $E_{\rm F}$ is the Fermi energy, $k_{\rm F}$ the Fermi momentum, ${\rm q}=|\bs Q|/2=\pi/\sqrt 2 a$,
and having assumed a quadratic dispersion for the large circular FS --
which, we emphasize again, is still physically the only Fermi surface present, as the boson $\phi$ has not condensed and remains a dynamical field.
In the above, as long as $\Delta(\bs k,t)$ varies over momentum scales (later denoted $\Pi_x,\Pi_y$) that are much larger than $ \delta \! k_x,  \delta \! k_y$,
one need not specify its $\bs k$ dependence as its effect on Eqs.\eqref{eq:1089a}, \eqref{eq:1089b} is negligible.
The ratio of parameters $\delta \! k_x , \delta \! k_y$ defines the asymptotic geometry of the crossing (i.e.\ the slope of dashed blue lines in Fig.\ref{fig:gap_geometry}),
meanwhile their product is the momentum-space area of that region near hot spots
where the electronic semiclassical dynamics is strongly affected by coupling to the boson (i.e.\ the dashed pink rectangle in Fig.\ref{fig:gap_geometry}).
The latter quantity,
\begin{align}
  \label{eq:109}
 l_B^{-2}\, \tilde \mu(t) &= \tfrac 1 2 \delta \! k_x   \delta \! k_y,
\end{align}
will be useful in the following. 
Here we stress that it involves $\Delta(t)$, therefore it is a fluctuating quantity: not only is it time-dependent within a given $\phi_0$ sector,
but physical observables only involve its averaged value over bosonic configurations (i.e., at the saddle-point, over $\phi_0$).
The momentum dependence $\Delta(\bs k,t)$ is kept implicit at this stage, however it will play a role in the time-dependent problem.
We note, for future use, the more explicit form
\begin{align}
  \label{eq:99}
   l_B^{-2}\, \tilde \mu(\bs k,t) &= (1/v_{\rm F}^{2})\,\sigma  \left | \Delta (\bs k,t)  \right |^2 ,
\end{align}
where $v_{\rm F}=k_{\rm F}/m_{\text \Circle}$ is the Fermi velocity, $ \left | \Delta (\bs k,t)  \right |^2 $ is given in Eq.\eqref{eq:483},
and $\sigma$, Eq.\eqref{eq:16},
is a dimensionless factor that only depends on the broad geometrical features of Fig.\ref{fig:gap_geometry}.

\subsection{In a magnetic field: effective tunneling problem}
\label{sec:effect-tunn-probl}

When a magnetic field is switched on, the theory near the hot spot pair can be formulated as a tunneling problem.
For a crossing of two Fermi surface portions that are both electron-like (or both hole-like), the resulting problem has a two-band structure,
while for tunneling between one electron-like and one hole-like Fermi pocket the problem is effectively single-band.
This determines the form of the scattering matrix $\mathbb S$ at the junction, which in our case (the former) will be given by Eq.\eqref{eq:116} below.
All physical quantities in the scattering matrix are determined as a function of a single parameter, $\zeta$,
which in the \emph{static} case (recovered in the limit $\Omega \rightarrow 0$) is identical to $\tilde \mu(t)\equiv\mu$ (that is $t$-independent),
but crucially in the \emph{dynamical} case can be strongly enhanced.

Apart from the form of $\mathbb S(\zeta)$, the parameter $\zeta$ itself can be determined independently from the matrix structure of the tunneling problem,
be it one-band or two-band -- see Appendix \ref{sec:mapping-weber-eqns} for details. Therefore, for notational simplicity in the following we adopt a single-band description.
First we consider the static case, in a magnetic field $B$, using the Landau gauge $k_x \rightarrow k_x - y/l_B^2$ with $l_B=({\sf e}B)^{-1/2}$ the magnetic length.
The Schrödinger equation $\bar H_{\rm eff} \psi = \mc E \psi$, at the particular energy value $\mc E= \mu\, l_B^{-2} (m_xm_y)^{-1/2}$, upon changing variables
\begin{subequations}
  \label{eq:99}
\begin{align}
   \label{eq:99a}
  k_x \quad &\mapsto \quad l_B^{-1}\sqrt{ \delta \! k_x/ \delta  \! k_y}\, (m \kappa)^{-1/4}\, i\partial_z ,\\
  \label{eq:99b}
  k_y \quad &\mapsto \quad l_B^{-1}\sqrt{ \delta  \! k_y/ \delta \!  k_x}\, (m \kappa)^{+1/4} \, z ,
\end{align}
\end{subequations}
and $\tilde \psi_E = e^{-ik_xk_y}\psi_E $, becomes
\begin{align}
  \label{eq:90}
    \left [ - \tfrac 1 {2m} \partial^2_{zz} - \frac{\kappa}{2} z^2 -E \right ] \, \tilde \psi = 0 ,
\end{align}
where $m$ and $\kappa$ are arbitrary parameters and one defined $E = - \tfrac 1 2 (\kappa/m)^{1/2} \mu$. 
Thus, the problem of electron dynamics at a pair of hot spots has been rewritten as a tunneling problem through an inverted parabola \cite{blount1962bloch}.

Crucially, nothing in the above is specific to the static problem, and the problem with a time-dependent $\mu \rightarrow \tilde \mu(t)$
takes exactly the same form Eq.\eqref{eq:90} upon replacing $E \rightarrow - \tfrac 1 2 (\kappa/m)^{1/2}\tilde \mu(t)$.
Differences appear in the solution methods: while solving the static problem can be done exactly in terms of asymptotics of special functions (see Ref.\cite{berry1972semiclassical} for a review),
solving the time-dependent problem can only (as far as we know) be done perturbatively, e.g.\ by Floquet-type expansions \cite{GRIFONI1998229}
or in the classical action picture which we will discuss in Sec.\ref{sec:pert-analyt-solut}.

Explicitly, the problem we endeavor to solve is now the following:
\begin{align}
  \label{eq:100}
   \left [ - \tfrac 1 {2m} \partial^2_{zz} + V(z,i\partial_z,t) -E_0 \right ] \, \tilde \psi = 0 ,
\end{align}
where $E_0 = - \tfrac 1 2 (\kappa/m)^{1/2} l_B^{2}  v_{\rm F}^{-2}\sigma \Delta_0^2$ is the (uniform, constant) energy,
with $\Delta_0^2=\tfrac 1 {2} \left | g(\bs 0)\right |^2\, \phi_0^* \phi_0$,
and all the momentum and time dependence of the problem is now contained in
\begin{align}
  \label{eq:101}
  V(z,i\partial_z,t) &= -\frac{\kappa}{2} z^2 +
                    \tfrac 1 2  l_B^{2}v_{\rm F}^{-2}\sigma  (\kappa/m)^{1/2} \tfrac 1 {2} \phi_0^* \phi_0 \times \\
  &\qquad \times \left | g(\bs k)\right |^2_{\bs k = \left [\begin{smallmatrix} k_x(i\partial_z) \\ k_y(z) \end{smallmatrix} \right ]}  \varepsilon \cos\left ( 2\Omega t - 2 \alpha_0 \right ) ,  \nonumber
\end{align}
where $k_x(i\partial_z),k_y(z)  $ are given by Eqs.\eqref{eq:99a},\eqref{eq:99b}.
Here $\varepsilon$ is formally a small parameter, which will govern the perturbative expansion of the tunneling dynamics,
and ultimately will be taken to be $\varepsilon \rightarrow 1$.

Notice that in the change of variables Eqs.\eqref{eq:99} we introduced the unit parameters $m,\kappa$ that make Eq.\eqref{eq:90} look intuitive, as they identify as a mass and a spring constant.
Such scales do not appear in the original (static) problem, i.e.\ they are basically a ``gauge'' freedom, so any physical result (in the static case) should be independent of them -- which we will check below.
Meanwhile, the time-dependent problem sets a unit of time, and we will see below that the value of $\kappa/m$ must be fixed in that case, in order for Eq.\eqref{eq:100} to describe the physical hot spot problem.
Even in the time-dependent problem, $\kappa m$ remains a ``gauge'' redundancy, i.e.\ no physical result depends on it.

\subsection{Perturbative analytical solution and calculation of the scattering matrix}
\label{sec:pert-analyt-solut}

We now formulate the tunneling problem using the classical action
\begin{align}
  \label{eq:146}
  \mc S(E_0) = \int\text d t \left ( p \dot z -  \tfrac 1 {2m}  p^2 - V(z,p ,t) + E_0 \right ),
\end{align}
where the time integral is over the Hamilton-Jacobi contour (see details in Appendix \ref{sec:acti-integr-form}), and we identified $p\equiv i\partial_z$.
We may now employ the tools of semiclassical tunneling through a time-dependent barrier \cite{ivlev-melnikov,kamenev2011field},
with the notable difference that $V(z,p ,t)$ is not a \emph{potential} strictly speaking, as it depends on position $z$ but also on momentum $p$.
We are now in a position to evaluate the tunneling parameter $\zeta$, that will ultimately determine the full scattering matrix $\mathbb S(\zeta)$, perturbatively in $\varepsilon$.
Namely, from $\mc S = \mc S_0 + \varepsilon \mc S_1 + \dots$ one obtains $\zeta = \zeta_0 + \varepsilon \zeta_1 + \dots$.

To the zeroth order, one is back to the static problem. Evaluating the tunneling coefficient from the action integral Eq.\eqref{eq:146} is straightforward, and yields \cite{KANE1960181}
\begin{align}
  \label{eq:108}
  \zeta_0 = \frac 1 {i\pi} \mc S_0(E_0) = l_B^{2} v_{\rm F}^{-2}\sigma\Delta_0^2  .
\end{align}

To the first order, the precise form of the time-dependent barrier plays a role, in particular one needs to specify  $\left | g(\bs k)\right |^2$,
which we recall captures how the effective electron-boson interaction peaks at the hot spots.
For concreteness and analytical simplicity, we now assume
\begin{align}
  \label{eq:73}
  \left | g(\bs k)\right |^2 & = \frac{\bar g^2}{1 + (k_x/\Pi_x)^2+ (k_y/\Pi_y)^2},
\end{align}
where $\Pi_x,\Pi_y$ are parameters that set the size of the hot-spot coupled region and $\bar g^2=|g(\bs 0)|^2$ sets the strength of the coupling.
Again using Eqs.\eqref{eq:99} and identifying $p\equiv i\partial_z$, equivalently this means that Eq.\eqref{eq:101} reads
\begin{align}
  \label{eq:110}
  V(z,p,t) &= - \frac{\kappa}{2} z^2
             + \varepsilon E_0 \,  \frac{ \cos\left ( 2\Omega t - 2 \alpha_0 \right )} {1+ (z/\xi)^2 - (p /m v)^2}  ,
\end{align}
where $ \xi^{-1}, (mv)^{-1}  $ are defined in Eqs.\eqref{eq:36a}-\eqref{eq:36b}.

The simplicity of this form of $ V(z,p,t) $ allows to evaluate analytically the action integral to the order $\varepsilon^1$: for details see Appendix \ref{sec:time-depend-oscill}.
The result is
\begin{align}
  \label{eq:111}
\varepsilon \zeta_1 & = \frac \varepsilon {i\pi} \mc S_1(E_0) \nonumber\\
  &= \varepsilon \frac{\sigma\Delta_0^2 \,  l_B^{2}v_{\rm F}^{-2}  \,\cos(\lambda) \, \cosh(\Omega t_0)}{\sqrt {\left ( 1 +   (z_\star/ v t_0)^2 \right ) \left ( 1+  (z_\star/\xi)^2  \right )}} ,
\end{align}
where $\lambda$ is a rather complicated phase that ultimately will prove unimportant,
and $z_\star = \sqrt{-2E_0/\kappa}$ and $t_0 =  \pi \sqrt{m/\kappa} $  are characteristic quantities of the time-independent tunneling problem.
At this stage, it is easy to check that the quantities $z_\star/ v t_0$ and $z_\star/\xi$ appearing in the denominator are independent of $\kappa,m$.
Meanwhile, and crucially, $\cosh(\Omega t_0)$ is not: in the time-dependent problem,
the unit of time $t_0$ appears in physical quantities and needs to be fixed on physical grounds.
One can use the fact that $t_0$, an imaginary time, identifies as the real time it takes an electron to cross the same region in the inverted potential:
this simply means crossing the pink hot spot region in Fig.\ref{fig:gap_geometry}, traveling a distance $2\delta\!k = 2\sqrt{\delta\!k_x^2+\delta\!k_y^2}$ in momentum space,
for an electron obeying semiclassical dynamics. Since such an electron orbits around the large FS with the cyclotron period $2\pi/\omega_{\rm c}$
(where $\omega_{\rm c}={\sf e}B/m_{\text \Circle}$), the crossing time is $t_0 = (2\delta\!k /2\pi k_{\rm F})\times (2\pi/\omega_{\rm c})$.
This fixes the ratio
\begin{align}
  \label{eq:112}
 (m/\kappa) = (t_0/\pi)^2 = (1/\pi)^2 E_{\rm F}^{-2} \Delta_0^2\,\sigma^2 /\omega_{\rm c}^2 ,
\end{align}
so that all the quantities appearing in Eq.\eqref{eq:111} are uniquely determined in terms of the physical parameters of the system
and of the bosonic amplitudes $\phi_0,\phi_0^*$ in the given boson sector one is working in.

The limit $\bar g \rightarrow 0$ provides a good sanity check, implying $\Delta_0^2\rightarrow 0$, Eqs.\eqref{eq:108}, \eqref{eq:111} yield $\zeta \rightarrow 0$:
when electrons of the large FS are not coupled to bosons, no tunneling happens between the hot spots. 
In the limit $l_B\rightarrow 0$, similarly, $\zeta \rightarrow 0$: when electron wavepackets are spread in momentum space (see Fig.\ref{fig:gap_geometry})
with size $l_B^{-1}$ much larger than the width of the boson-induced avoided-crossing, tunneling between the hot spots is ineffective.
In the limit $k_{\rm F} = \rm q$ of nesting, $\sigma \rightarrow \infty$ and so $\zeta \rightarrow \infty$: boson-induced tunneling is then very effective.

\section{Orbit quantization and physical consequences}
\label{sec:orbital-quantiz-phys-conseq}

\subsection{Formula for the scattering matrix}
\label{sec:form-scatt-matr}

We still work with $\zeta \approx \zeta_0+\zeta_1$ (having restored  $\varepsilon \rightarrow 1$) within a given bosonic sector,
i.e. for fixed values of $\phi_0^*,\phi_0$.
At each tunneling junction (i.e.\ hot spots pair), the scattering matrix $\mathbb S(\zeta)$ relates incoming and outgoing wavefunction amplitudes $a_{L/R}^\pm$
(see schematic in Fig.\ref{fig:gap_geometry}) according to the tunneling formula
\begin{align}
  \label{eq:115}
  \begin{pmatrix}
   a_R^-\\a_L^-
  \end{pmatrix} = {\mathbb S} \begin{pmatrix}
   a_R^+\\a_L^+
  \end{pmatrix} ,
\end{align}
where ``$+$'' means incoming, ``$-$'' means outgoing, and $R/L$ means that the semiclassical trajectory is deviating to the right/left at the avoided crossing.
The scattering matrix for the problem defined by Eq.\eqref{eq:90} (which we generalize here to its time-dependent counterpart) is known
\cite{azbel1961quasiclassical,azbel1964energy,davis1967landau,slutskin1968dynamics,berry1972semiclassical}, and reads
\begin{align}
  \label{eq:116}
  {\mathbb S}(\zeta) &=
  \begin{pmatrix}
    \sqrt{1-\rho^2}e^{i\omega} & -e^{i\delta}\rho \\ e^{-i\delta}\rho & \sqrt{1-\rho^2}e^{-i\omega} 
  \end{pmatrix} .
\end{align}
Here $\rho=e^{-\pi \zeta}$ is the tunneling real amplitude, 
and $\omega = \zeta - \zeta \ln \zeta + \pi/4 + {\rm arg}[{\sf \Gamma}(i\zeta)]$, where $\sf \Gamma$ is Euler's gamma-function,
with limits $\omega \overset{\zeta \rightarrow 0}\rightarrow  -\pi/4$ and $\omega \overset{\zeta \rightarrow \infty }\rightarrow  0$.
In general $\delta$ is a relative phase between electronic wavefunctions in different bands or pockets, that can be fixed arbitrarily
since all physical results are independent of this choice. Here, it is still possible to define effectively two ``bands'' (see Fig.\ref{fig:gap_geometry}b)
and we fix $\delta=-\pi/2$ so as to match the intuition from quantum-mechanical tunneling \cite{LL3} that
each tunneling event comes with probability $\rho^2$ and a phase factor $i$.

\subsection{Thermal average over bosonic fluctuations}
\label{sec:thermal-average-over}

At this stage, $\zeta$ still depends on the bosonic amplitudes $\phi_0,\phi_0^*$.
Tunneling characteristic quantities now need to be averaged over bosonic fluctuations, which at the saddle-point means over $\phi_0,\phi_0^*$.
Equivalently this is averaging over the phase $\alpha_0$ (which can be done at zero temperature) and over the bosonic number $\nu_0=\phi_0^*\phi_0$ at a finite temperature $T$.
Details are provided in Appendix \ref{sec:phys-observ-aver}.

Explicitly, we perform the thermal average over bosonic fields in the following way:
\begin{align}
  \label{eq:117}
  \langle \bullet \rangle = \frac 1 {\mc Z} \int \frac{\text d \alpha_0}{2\pi}\sum_{\nu_0=0}^\infty e^{-\nu_0\Omega/T} \bullet ,
\end{align}
with $\mc Z=\left ( 1 - e^{-\Omega/T} \right )^{-1}$ the bosonic partition function.
In particular, one obtains the tunneling probability $1-\langle\rho^2\rangle$, 
\begin{align}
  \label{eq:102}
 \langle\rho^2\rangle =  \left  \langle  e^{-2\pi \zeta} \right \rangle
  &= \frac 1 {\mc Z} \sum_{\nu_0=0}^\infty e^{-\nu_0 \Omega / T}e^{-2\pi  l_B^{2}v_{\rm F}^{-2}\sigma \Delta_0^2 } \times\\
   &\quad \times {\sf I}_0\left ( \frac{ 2\pi  l_B^{2} v_{\rm F}^{-2}\sigma \Delta_0^2 \, \cosh(\Omega t_0)}
                           {\sqrt {\left ( 1 +   (z_\star/v t_0)^2 \right ) \left ( 1+  (z_\star/\xi)^2  \right )}} \right ) , \nonumber
\end{align}
where ${\sf I}_0$ is the zeroth modified Bessel function.
We also recall that $\Delta_0^2=\tfrac 1 {2} \bar g^2 \nu_0$ and that in the second line of Eq.\eqref{eq:102},
$t_0$, $z_\star/v t_0$ and $z_\star/\xi$ all depend on $\Delta_0$.
In the limit $T \rightarrow 0$, Eq.\eqref{eq:102} gives $\langle \rho^2 \rangle \rightarrow 1$: when bosonic states are not populated, no tunneling happens.

At this stage, a discussion of the validity of our first-order $\mc S = \mc S_0 + \varepsilon \mc S_1 + \dots$ perturbative expansion developed in Sec.\ref{sec:pert-analyt-solut},
and of the convergence of the sum in Eq.\eqref{eq:102}, is in order: see Appendix \ref{sec:techn-comm-conv} for details.

\subsection{Connection formula}
\label{sec:connection-formula}

\begin{figure}[htbp]
  \begin{center}
    \includegraphics[width=1 \columnwidth]{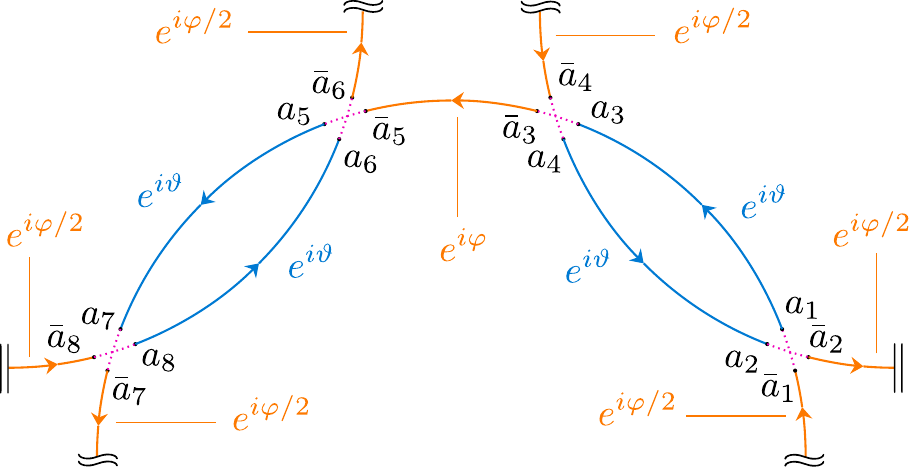}
    \caption{Summary of wavefunction amplitudes at several points of the FS, and of phase factors accumulated along semiclassical paths.
      The red and blue semicircles belong to a single large circular FS. The ``$\approx$'', ``\text{\reflectbox{$\approx$}}'' and ``$\parallel$''
      indicate (wavefunction) periodicity along the physical FS. Dotted lines are in the hot spot regions.
    Factors $e^{i\vartheta},e^{i\varphi/2}$ are the phases acquired by electrons along the corresponding semiclassical paths.}
\label{fig:schema}
\end{center}
\end{figure}

At the core of the semiclassical picture is the existence of unambiguous electronic trajectories and energy levels, and their correspondence.
Clearly, the right panel of Fig.\ref{fig:advertisement} is only a cartoon, but our semiclassical treatment of dynamical magnetic breakdown allows
such an unambiguous characterization, determined through a \emph{connection formula} \cite{azbel1961quasiclassical,pippard1962quantization,azbel1964energy,pippard1964quantization,
  davis1967landau,slutskin1968dynamics}.

We briefly come back to the picture before thermal averaging, where $\zeta$ is fixed
and tunneling at the hot spots is fully described by $\mathbb S(\zeta)$, Eq.\eqref{eq:116}.
Then, electronic wavefunctions are constrained to be eigenmodes of the semiclassical dynamics,
whereby electrons travel along the semiclassical paths shown in Fig.\ref{fig:schema} acquiring phases $\vartheta$ or $\varphi$
(including Maslov phases \cite{keller1958corrected,roth1966semiclassical}).
Their amplitudes $a_i,\bar a_i (i=1..8)$ are connected at the hot spot regions through relations like Eq.\eqref{eq:115}.

Thus, looking for ``standing waves'' of the ``circuit'' depicted in Fig.\ref{fig:schema}, one can obtain a connection formula,
that determines precisely the semiclassical phases $\vartheta,\varphi$ and equivalently, via Bohr-Sommerfeld quantization, the semiclassical energy levels.
Details of the derivation are provided in Appendix \ref{sec:conn-form-cupr}.
The resulting connection formula is
\begin{align}
  \label{eq:1066}
  0 & =\mc F_{\rho,\omega}(\vartheta,\varphi) ,
\end{align}
where
\begin{align}
  \label{eq:106}
 \mc F_{\rho,\omega}(\vartheta,\varphi)  & \equiv  \sin(2\vartheta + 2\varphi) - 2(1-\rho^2)\sin(2\varphi-2\omega) \nonumber \\
                                         & \; + 4 \rho^2 \sqrt{1-\rho^2} \sin(\vartheta+\omega) \nonumber \\
  &\; - (1-\rho^2)^2 \sin(2\vartheta - 2\varphi + 2\omega).  
\end{align}

This fixes, as a function of $\zeta$ (through $\rho,\omega$), the energies and trajectories in momentum space of the semiclassical orbits,
from which the amplitude of magneto-oscillations will be obtained (see Sec.\ref{sec:qos-their-temp} below).
One can check that this formula matches intuition in limit cases -- separating the Maslov phases
from other semiclassical phases via $\varphi =\varphi_{\rm cl} - \pi/4, \vartheta = \vartheta_{\rm cl}  - \pi/2$.
\begin{itemize}
\item[-] In the limit $\zeta \rightarrow \infty$, the connection formula reads $\big ( \cos(2\vartheta) - 1 \big )  \sin(2\varphi) = 0$,
  with solutions $4\varphi_{\rm cl} \equiv \pi \,[2\pi] \;\text{ or }\; 2\vartheta_{\rm cl} \equiv \pi \,[2\pi] .$
  This quantizes the reconstructed orbits, enclosing areas $S_{( \! )}$ and $S_{\text \Square}$.
 \item[-]  In the limit $\zeta \rightarrow 0$, the connection formula reads $\sin(2\vartheta + 2\varphi) =0$,
   with solutions $4\varphi_{\rm cl}+4\vartheta_{\rm cl} \equiv \pi \,[2\pi].$
   This quantizes the orbits around the original large FS, enclosing area $S_{\text \Circle}$.
\end{itemize}

Because $\zeta$ in the present case is ultimately a fluctuating quantity, the quantization condition Eq.\eqref{eq:106}
will need to be thermally averaged over bosonic fields $\phi_0,\phi_0^*$.
In principle, this will generate a thermal damping of quantum oscillations, following Shoenberg's phase smearing arguments \cite{shoenberg1984magnetic},
if thermal fluctuations of the phases $\vartheta,\varphi$ solutions of Eq.\eqref{eq:1066} become of the order of $\pi$.
It is possible, however, to show that no significant damping is generated this way in general: see Appendix \ref{sec:techn-deta-therm}.

\subsection{QOs and their temperature dependence}
\label{sec:qos-their-temp}

In the absence of any electron-boson scattering, the only magneto-oscillation corresponds to the quantization of orbits around the full circular FS,
with a large frequency $S_{\text \Circle}/\sf e$ ($\sf e$ the electron charge), and a temperature dependence obeying the standard Lifshitz-Kosevich law
with the effective mass $m_{\text \Circle} = k_{\rm F}^2/2E_{\rm F}$.
Once electron-boson scattering is switched on, dynamical magnetic breakdown orbits become available,
in particular those with frequencies $S_{( \! )}/\sf e$ and $S_{\text \Square}/\sf e$,
where the areas $S_{( \! )},S_{\text \Square}$ are identified in Fig.\ref{fig:advertisement} and expressed in Eqs.\eqref{eq:68a}-\eqref{eq:68b}.

\begin{figure}[htbp]
  \begin{center}
    \includegraphics[width=\columnwidth]{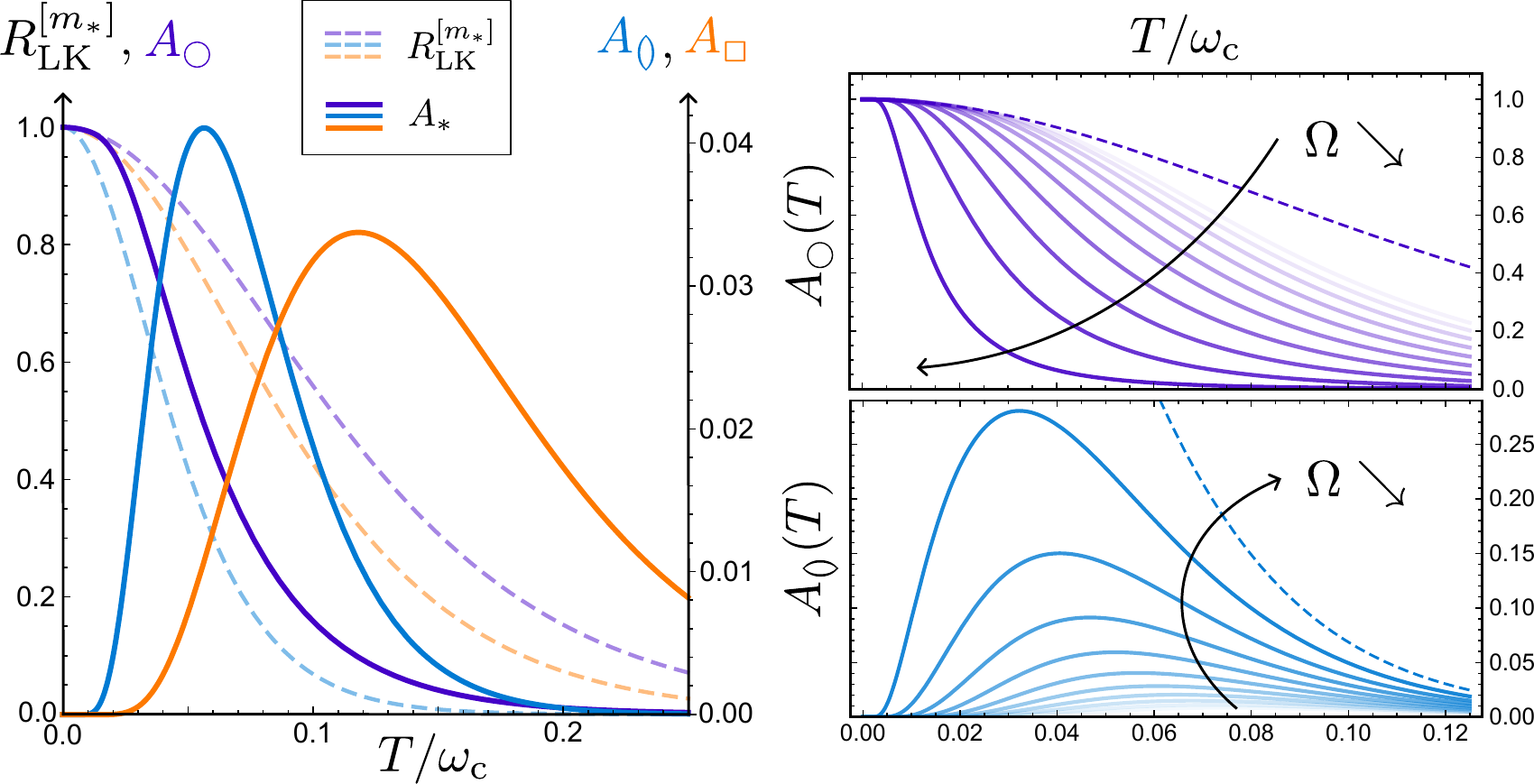}
    \caption{Left: the temperature dependence of the QO amplitudes $ A_\ast(T)  $ for $\ast= \text {\small (} \! \text {\small )} , \text{\Square, \Circle}$
      does not obey the Lifshitz-Kosevich dependence, i.e.  $R_{\rm LK}^{[m_\ast]}(T)$ defined from Eq. \eqref{eq:69}. Right: evolution of the curves $A_{\text \Circle}(T), A_{(\!)}(T)$
      for decreasing $\Omega \in [0.02,0.2].$ The dashed curves are in the limits $\Omega \rightarrow \infty$ (for $A_{\text \Circle}$) and $\Omega \rightarrow 0$ (for $A_{(\!)}$),
      and equate the corresponding dashed curves in the left panel. Note that here $\Omega,m_\star$ are treated as independent from $B,T$.
    The parameter values used to generate the figure are given in Table \ref{tab:parameter-values}.}
\label{fig:plots}
\end{center}
\end{figure}

Their amplitude does not reduce to the corresponding Lifshitz-Kosevich law, with effective mass
\begin{align}
  \label{eq:70}
  m_\ast = (2\pi)^{-1}\, \partial S_{\ast}/\partial E_{\rm F}
\end{align}
for $\ast = \text {\small (} \! \text {\small )} , \text{\Square, \Circle}$.
Instead this must be multiplied with a probability weight given by the thermal-averaged Chambers-Falicov formula \cite{chambers1966magnetic,falicov1966theory},
$P_{( \! )} = \langle 1-\rho^2 \rangle$, $P_{\text \Square} = \langle (1-\rho^2)^2 \rangle$, and $P_{\text \Circle} = \langle \rho^8 \rangle$
for the orbits enclosing area $S_{( \! )}, S_{\text \Square}$ and $S_{\text \Circle}$ respectively.
Crucially, because such an average (for instance $\langle \rho^2 \rangle$ as given by Eq.\eqref{eq:102}) depends on temperature,
magneto-oscillations have a non-Lifshitz-Kosevich behavior. In particular,
the dynamical magnetic breakdown frequencies $S_{( \! )}/\sf e$ and $S_{\text \Square}/\sf e$ display a thermal activation behavior at low temperature.
The (normalized) amplitudes of magneto-oscillations for each frequency,
\begin{subequations}
 \label{eq:69}
\begin{align}
  A_\ast(T)   &= R_{\rm LK}^{[m_\ast]}(T)\, P_\ast(T) ,\\
 R_{\rm LK}^{[m_\ast]}(T) & := \frac{2\pi^2 l_B^2 T m_\ast}{\sinh(2\pi^2 l_B^2 T m_\ast)},
\end{align}
\end{subequations}
for $\ast= \text {\small (} \! \text {\small )} , \text{\Square, \Circle}$, are plotted in the left panel of Fig.\ref{fig:plots}.
Note that Eq.\eqref{eq:69} omits the standard Dingle factors \cite{shoenberg1984magnetic} which are not our concern here.

The key new finding of our work is that new semiclassical orbits are dynamically allowed by a boson retaining a finite energy gap $\Omega$, thus, before it condenses leading to long-range density order. This dynamical magnetic breakdown effect is manifest through the QO amplitude dependence on $\Omega$.
A smaller bosonic energy $\Omega$ increases the thermal population of bosonic states, thereby enhancing the scattering between hot spot pairs.
This reduces the amplitude of QOs at the original frequency $S_{\text \Circle}/\sf e$, 
while increasing the amplitude of scattering-allowed orbits with corresponding frequencies $S_{(\!)}/{\sf e}, S_{\text \Square}/{\sf e}$ among others.
This behavior is illustrated in the right panel of Fig.\ref{fig:plots}, focusing on $A_{\text \Circle}(T)$ and $A_{( \! )}(T)$.

Note that here, the only effect of temperature we consider is that it determines the thermal population of bosonic states. We do not include in Eq.\eqref{eq:69} its effects on other parameters of the model itself, such as $\bar g$ or $\Omega$. In particular, at lower $T$ one expects an ordering transition corresponding to the boson condensing, not included here -- it can also be that several orders are competing, and that another boson than Eq.\eqref{eq:74}'s $\phi$ actually condenses first. For a realistic modeling of concrete systems, such effects should be taken into account, along with other parameter dependencies, especially when varying the magnetic field (which can also trigger phase transitions).

\begin{table}[htbp]
  \centering
  \begin{tabular}{c|c|c|c|c|c|c|c}
    \hline\hline
     \; $a$ \;  & \;  $k_{\rm F}$  \; & \;  $l_B$ \; &\;   $m_{\text \Circle}$ \;    & \;  $\Omega$ \; &  \; $\bar g$\;  & \;  $\Pi_x$ \; &  \; $\Pi_y$ \; \\   \hline
   $1.0$ & $2.75$ & $10$  & $0.01$  & $0.1$ & $10$  & $0.2$  &  $0.2$  \\
    \hline\hline
  \end{tabular}
  \caption{Parameter values used for the numerical evaluations plotted in Fig.\ref{fig:plots}, unless stated otherwise.}
  \label{tab:parameter-values}
\end{table}

\subsection{Discussion}
\label{sec:discussion}

The extra tunneling probability factors $P_*(T)$ appearing in Eq.\eqref{eq:69}, which causes the deviations from Lifshitz-Kosevich behavior visible in Fig.\ref{fig:plots}, is equivalent in the high-temperature limit to a mass renormalization effect. Namely, stronger bosonic fluctuations appear as increasing the mass of the original ($S_{\text \Circle}$) orbit, and as decreasing the mass of the reconstructed ($S_{(\!)}, S_{\text \Square}$) orbits. Such an effect of effective mass enhancement when approaching criticality was indeed found experimentally in underdoped YBCO \cite{sebastian2010metal}. 
Of course, for a quantitative comparison the interaction effects away from the hot spots, which renormalize the effective mass and the Dingle factor, should also be included. Besides, the temperature dependence we find in Fig.\ref{fig:plots} only appears as a mass renormalization at higher temperatures and it would be worthwhile to look for the predicted deviations at lower temperatures.

From a theory standpoint, the fact that electron-boson interactions can yield nontrivial temperature dependence is unusual. Indeed, early works on the electron-phonon coupling \cite{fowler1965electron, engelsberg1970influence} showed that the temperature behavior of Onsager QOs, when interactions are included perturbatively, remains of Lifshitz-Kosevich type, only with a renormalized mass. This (Fowler-Prange) theorem is linked to a fundamental Fermi-liquid property, the ``first Matsubara frequency rule'' \cite{martin2003quantum} -- deviations from this behavior are found in non-Fermi liquids \cite{nosov2024entropy}, which is not the case we consider here.
Instead, the magnetic breakdown scenario we discuss here relies on the formation of new semiclassical electron trajectories. These correspond to a global rearrangement of Landau levels, that presumably cannot be achieved in a perturbative treatment, as it results from an all-to-all hybridization of the Landau levels. Our semiclassical treatment, albeit limited to coherent enough bosonic fluctuations and small enough hot spot regions without any interaction effects outside, allows to overcome this difficulty. The Fowler-Prange theorem, on the other hand, only applies to the original semiclassical orbits and perturbations thereof due to interactions, therefore it misses non-perturbative effects such as the appearance of new semiclassical paths. 

We stress that the emergence of new well-defined semiclassical orbits relies on the existence of sharp hot spots. Were the interacting regions more extended, with broad momentum-dependent features, no sharply defined semiclassical paths could be defined except that enclosing $S_{\text \Circle}$. In other words, the semiclassical phases $\vartheta,\varphi$ themselves would have to be treated as fluctuating variables (see in that respect App.\ref{sec:techn-deta-therm}), and since these determine reconstructed Landau level energies via Bohr-Sommerfeld quantization and Eq.\eqref{eq:1066}, this would necessarily damp out the amplitudes of reconstructed orbits per Shoenberg's argument \cite{shoenberg1984magnetic}.

\section{Conclusion}
\label{sec:conclusion}

In this paper, we have shown that when a large Fermi surface is coupled to bosonic near-critical fluctuations, with interactions concentrated near a set of hot spots, then ``reconstructed'' QO frequencies appear even without assuming condensation of the boson, i.e.\ without a non-zero order parameter of an underlying long-range density wave order. To show this, we used a semiclassical saddle-point approach, treating the electron-boson interaction as an effective time-dependent tunneling event at the hot spots. Because the bosonic field in our theory remains fluctuating, the tunneling probabilities (that determine the magnetic breakdown QO amplitudes) depend on the thermal activation of bosonic excitations, and acquire unusual temperature dependences. Interestingly, also the original large frequency orbit aquires a non-LK temperature decay because of hot-spot scattering. 

Our predictions, both of the possibility of such ``reconstructed'' QOs in the normal metallic state and of their non-Lifshitz-Kosevich behavior, rely on the presence of sharply defined hot spots where electron-boson interactions are concentrated. The semiclassical method we used, albeit approximate, allows one to address the problem of QOs in strongly interacting settings, including phenomena such as orbit reconstruction that are non-perturbative in the basis of Landau levels. As a natural extension beyond QOs, it would be interesting to investigate signatures of Berry phases, e.g.~the AHE (as a FS property \cite{haldane2004berry}), and how they differ in the reconstruction vs the dynamical magnetic breakdown scenario.

On the experimental side, the observation of new QO frequencies have been taken as a sign of reconstructed FS from long-range order. Our results show that the reconstructed frequencies may appear prior to the transition and can be distinguished via their distinct temperature dependence. An observation of a suppressed amplitude, with an activated behavior set by the boson gap (before the latter closes at a thermal phase transition), would help to refine the phase diagram of quantum critical metals and quantify their scattering properties. For example, a controlled tuning of the boson gap, e.g.~by pressure, would allow to extract the electron-boson scattering strength. Beyond parent phases of high-temperature superconductors, nesting-driven density wave transitions have also been discussed for heavy fermion systems like URu$_2$Si$_2$~\cite{hassinger2010similarity,shishido2009possible}. There, the reduced energy (temperature) scales and tunability of the phase diagram with non-thermal knobs like pressure should be ideally suited for a controlled study of scattering-induced QOs.

Looking forward, our semiclassical treatment gives access to other magnetic field dependent physical properties, that also rely on semiclassical orbits, such as weak localization effects. In particular, because hot spot scattering enables new paths that break spatial symmetries (ultimately resulting in highly anisotropic reconstructed Fermi pockets), signatures will be found in direction-dependent quantities, e.g.~angle-dependent magnetoresistance or the anomalous skin effect. 
Our description also extends to 3D models, so that the effects of an in-plane field component can also be investigated. There, the hot region becomes a one-dimensional manifold, that intersects with extremal electron orbits (defined for a given field orientation) at a discrete set of points. In general, despite the fact that these points indeed belong to an extremal orbit, their translation by the ordering vector \emph{not necessarily does}, and therefore QOs from such reconstructed but non-extremal orbits will be suppressed. However, for some specific field orientations, a large number of orbits qualify as extremal: this defines the Yamaji angle \cite{yamaji1989angle}. In such a situation, our theory again applies. Notably, there has been recent interest in the Yamaji effect in underdoped cuprates \cite{chan2025observation, zhao2025yamaji}, that could also be discussed in the magnetic breakdown framework we established here. Besides, it would also be interesting to investigate signatures of open trajectories, that have been predicted to appear in Shubnikov-de Haas oscillations \cite{krix2024quantum}, also in (strongly) interacting settings, possibly including Stueckelberg intereferences in the magnetoresistance.

Finally, we note that because Luttinger's theorem at a given doping $\sf p$ fixes the area $S_{\text \Circle}$, simply by geometry it also imposes the relation
\begin{align}
  \label{eq:82}
  4 S_{( \! )} - 2S_{\text \Square} = {\sf p} 
\end{align}
which together with Eq.\eqref{eq:68b} fixes the areas  $S_{( \! )}$ and $S_{\text \Square}$, regardless of whether the boson condenses i.e.\ even in the non-reconstructed scenario. In other words, the QOs arising from dynamical magnetic breakdown that we discussed here obey the same Luttinger counting as per FS reconstruction. Under which realistic circumstances QO frequencies violating Luttinger's count can appear is an exciting question for future research.

\section*{Acknowledgements}
\label{sec:acknowledgements}
We thank Valentin Leeb and Peng Rao for valuable discussions. We thank Philipp Moll for helpful comments on the manuscript. We acknowledge support from the Deutsche Forschungsgemeinschaft (DFG, German Research Foundation) under Germany’s Excellence Strategy (EXC–2111–390814868 and ct.qmat EXC-2147-390858490),
and DFG Grants No. KN1254/1-2, KN1254/2-1 TRR 360 - 492547816 and SFB 1143 (project-id 247310070), as well as the Munich Quantum Valley, which is supported by the Bavarian state government with funds from the Hightech Agenda Bayern Plus. JK thanks the Keck foundation for support.

\bibliography{library}

\appendix

\section*{Explicit expressions}
\label{sec:explicit-expressions}

Here we gather explicit expressions of quantities appearing in the main text:
\begin{align}
  \label{eq:16}
 \sigma &= 2 / \sqrt{ ({\rm q}/k_{\rm F})^2 \big ( 1 -   ({\rm q}/k_{\rm F})^2 \big )},\\
    \label{eq:36a}
    \xi^{-1} &= l_B^{-1}\,\Pi_y^{-1}(m\kappa)^{1/4}\left ( k_{\rm F}^2/{\rm q}^2 -1 \right )^{-1/4} , \\
     \label{eq:36b}
  (mv)^{-1} &= l_B^{-1} \,\Pi_x^{-1}(m\kappa)^{-1/4}\left ( k_{\rm F}^2/{\rm q}^2 -1 \right )^{1/4}  ,\\
  \label{eq:68a}
    S_{( \! )} &= 2k_{\rm F}^2 \arccos({\rm q}/k_{\rm F}) - 2 \sqrt{k_{\rm F}^2 {\rm q}^2- {\rm q}^4 }, \\
    \label{eq:68b}
S_{\text \Square} &= 2 \pi^2/a^2 - \pi k_{\rm F}^2 + 2  S_{( \! )}  .
\end{align}

\section{Perturbative derivation of the effective low-energy theory}
\label{sec:pert-deriv-effect}

\subsection{Renormalized interaction vertex}
\label{sec:renorm-inter-vert}

We start from the interaction Eq.\eqref{eq:28} in the main text,
and we assume that the boson $b_{\bs q} $ disperses around $\bs Q=(\pi/a,\pi/a)$.
Our goal here is to obain an effective low-energy theory, valid around the hot spot pairs.

The bosonic dispersion does not appear in the bare coupling $g^{(0)}_{\bs q}$ of Eq.\eqref{eq:28},
however it will affect the renormalized coupling, that can be obtained from vertex corrections to the interaction term.
After performing such corrections, the coupling becomes effectively energy-dependent, so in the following we will use combined notations
$k=(\bs k, ik_n)$ for fermions [with $k_n = (2n+1)\pi T$] and $q=(\bs q, iq_n)$ for bosons [with $q_n = 2n\pi T$].

\begin{figure}[htbp]
  \begin{center}
    \includegraphics[width=.8 \columnwidth]{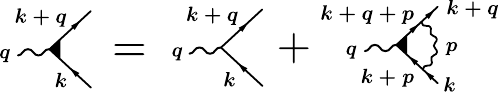}
    \caption{Diagrammatic summary of the vertex correction equation.}
\label{fig:vertex-eqn}
\end{center}
\end{figure}

The vertex correction equation (see e.g.\cite{migdal1958interaction}) is represented diagrammatically in Fig.\ref{fig:vertex-eqn},
and its expansion to the first perturbative order reads
\begin{align}
  \label{eq:15}
  \Gamma(\bs k,\bs q) &=  \Gamma^{(0)}(\bs k,\bs q) + \Gamma^{(1)}(\bs k,\bs q)  + \dots, 
\end{align}
\begin{widetext}
  where  $\Gamma^{(0)}(\bs k,\bs q) = g^{(0)}_{\bs q}$ is the bare electron-boson vertex and
\begin{align}
  \label{eq:72}
\Gamma^{(1)}(\bs k,\bs q)  &= - g^{(0)}_{\bs q}\int_{p}  |g^{(0)}_{\bs p} |^2
                                       \, G_0(k+p+q)\, G_0(k+p)\,D_0(p) \\
&= - g^{(0)}_{\bs q}\int_{\bs p}  |g^{(0)}_{\bs p} |^2 \sum_{\varsigma=\pm} n_{\rm B}(\varsigma \omega_{\bs p})
                               \,G_0(\bs k+\bs p+\bs q,ik_n+iq_n + \varsigma \omega_{\bs p} )\, G_0(\bs k+\bs p, ik_n+\varsigma \omega_{\bs p} ) .
\end{align}
Here we used the usual notations $G_0(k)=G_0(\bs k,ik_n)= 1/[ik_n - \epsilon_{\bs k}]$
and $D_0(p) = 1/[(ip_n)^2 - \omega_{\bs p}^2]$ for the bare fermionic and bosonic propagators,
and in going to the second line one performed the summation over bosonic $ip_n$.

\end{widetext}

Following e.g. \cite{abanov2003quantum} we now ask where the dominant contributions to Eq.\eqref{eq:72} come from,
in the low-energy limit where $ik_n\rightarrow 0, iq_n \rightarrow 0$.
Clearly, the integrand becomes large for vanishing denominators:
this means that the integral is dominated by those regions where $\epsilon_{\bs k+\bs q+\bs p} \approx \pm \omega_{\bs p} \approx \epsilon_{\bs k+\bs p}$.
Meanwhile, the factor $n_{\rm B}(\pm \omega_{\bs p})$ selects regions where the boson energy $\omega_{\bs p}$ is small (so $\bs p \approx \bs Q$),
which implies that the fermionic momenta $\bs k+\bs q+\bs p$ and $\bs k+\bs p$ sit close to the Fermi surface.

Here, we cannot \textit{a priori} assume anything about the external boson at $q$ since the bare coupling $g^{(0)}_{\bs q}$ does not impose any strong constraints \textit{a priori},
meanwhile we are focusing on external fermionic momenta $\bs k,\bs k+\bs q$ that also sit near the Fermi surface.
Together with the above, this implies that actually both $\bs k$ and $\bs k+\bs q$ are located near the hot spots (see Fig.\ref{fig:advertisement}).
In other words, the vertex correction $\Gamma^{(1)}(\bs k,\bs q)$ at low energy is dominated by fermionic momenta $\bs k$ near the hot spots
and by bosonic momenta $\bs q$ connecting two hot spots.
Further restricting our study to external bosonic lines with small energies $\omega_{\bs q}$, so that $\bs q \approx \bs Q$,
we can make a stronger statement: the low-energy vertex correction is dominated by fermionic momenta $\bs k,\bs k+\bs q$
sitting each near a hot spot belonging to the same pair.

\subsection{Effective low-energy Hamiltonian}
\label{sec:effective-low-energy}

Ultimately, the above perturbative derivation App.\ref{sec:renorm-inter-vert} serves as a plausibility argument for writing a low-energy hamiltonian that focuses on the hot spot physics.
We now assume a vertex-renormalized $H_{\rm int}^{(0)} \rightarrow H_{\rm int}^{\rm vr}$,
\begin{align}
  \label{eq:41}
H_{\rm int}^{\rm vr} &= \int_{\bs k, \bs q'} \tilde g(\bs k,\bs q')\; c^\dagger_{\bs{k+Q+q'}}  b_{\bs Q+\bs q'} c_{\bs k} + \rm h.c.,
\end{align}
where based on App.\ref{sec:renorm-inter-vert} we assume that
$ \tilde g(\bs k,\bs q')$ is peaked with respect of both its arguments,
around $\bs q'\approx \bs 0$ and crucially around momenta $\bs k$ that are close to a hot spot.

Because of the latter property, it is possible to write an effective theory focusing only on regions near the hot spots,
rewriting effectively $H_{\rm int}^{\rm vr}  \rightarrow H_{\rm int}^{\rm hs} $ where
\begin{align}
  \label{eq:34}
  H_{\rm int}^{\rm hs}  &= \int_{\bs q'} \sum_{\bs K \in {\rm HS}} \int'_{\bs k'} \tilde g(\bs K+\bs k',\bs q')\, b_{\bs{Q+q'}} \nonumber \\
  &\qquad \qquad \qquad \times c^\dagger_{\bs{K+k'+Q+q'}} \, c_{\bs{K+k'}} + \rm h.c. ,
\end{align}
where $\int'_{\bs k'} $ means that one integrates over small values of $\bs k'$ contained within a patch around the origin --
there is no need to specify explicitly a cutoff because the factor $\tilde g(\bs K+\bs k',\bs q')$ is peaked at small values of $\bs k'$ anyway.

The fact that $\tilde g(\bs k,\bs q')$ is peaked near $\bs q' \approx \bs 0$, as argued in App.\ref{sec:renorm-inter-vert},
was inherited through vertex renormalization from the assumption that the boson mode disperses sharply around momentum $\bs Q$.
At this stage, it is no longer strictly necessary to keep track of the momentum dependence of the bosonic field,
and for the sake of simplicity we perform the replacement
\begin{align}
  \label{eq:33}
\tilde g(\bs K+\bs k',\bs q') \,b_{\bs{Q+q'}} \quad \rightarrow \quad g(\bs K+\bs k') \, \phi \, \delta(\bs q') ,
\end{align}
where $g(\bs k) \equiv \int_{\bs q'} \tilde g(\bs k, \bs q')$ and $\phi$ is a single bosonic field that summarizes the bosonic fluctuations near $\bs q \approx \bs Q$.
Physically, this is reasonable provided that the bosons $b_{\bs q}$ at different momenta near $\bs q \approx \bs Q$ have similar dynamics,
which indeed is approximately satisfied in the case we consider here where the bosonic dispersion has an energy gap.
The boson's dynamics is now simply that of a single harmonic oscillator, $ H_\phi = \Omega \, \bar \phi \phi$,
with $\Omega$ an effective energy identifying approximately with the original boson gap.
As for its coupling to the fermions, one has thus obtained $H_{\rm int}^{\rm hs}  \rightarrow H_{\rm int}$ as given in Eq.\eqref{eq:30} in the main text.

\section{Saddle-point approximation}
\label{sec:saddle-point-appr-1}

Here we discuss in more detail the saddle-point approximation that allows to go from Eq.\eqref{eq:74} to Eq.\eqref{eq:93} in the main text.
Importantly, this goes beyond what is usually called the saddle-point approximation, where the bosonic field is assumed to condense to a finite value (thus becoming a mere number):
here the bosonic field retains some low-lying fluctuations. In other words, this is a more general saddle-point approximation
where the saddle-point manifold is still degenerate (and consists of time-dependent functions), so that the average of the boson field is still zero.

\begin{widetext}
Let us start from the full system's generating functional
\begin{align}
  \label{eq:1}
  \mc Z[ J_{\bar \phi} ,  J_{\phi} ,  J_{\bar \psi} ,  J_{\psi} ] &= \int \text D[\bar\phi,\phi] \text D[\bar \psi,\psi] \,
                                                                    e^{iS_{f}[\bar \psi, \psi] +iS_{b}[\bar \phi,\phi] +iS_{\rm int}[\bar \psi,\psi,\bar \phi,\phi]}
                                                                    \,  e^{i \int \left ( J_{\bar \phi} \bar \phi +J_{\phi} \phi +  J_{\bar \psi} \bar \psi +J_{\psi} \psi \right )} ,
\end{align}
where the free fermionic action $S_{f} = \int \text d t \,\mc L_{f}(t)$ is given in Eq.\eqref{eq:74},
the free bosonic action is $ S_{b}[\bar \phi,\phi] = \int \text d t \left (  i \bar \phi \partial_t \phi -  \Omega \,\bar \phi \phi \right ) $,
and $S_{\rm int}$ is the interaction part of the action (that need not be specified yet).

One can rewrite Eq.\eqref{eq:1} as
\begin{align}
  \label{eq:61}
  \mc Z[ J_{\bar \phi} ,  J_{\phi} ,  J_{\bar \psi} ,  J_{\psi} ]  &= \int \text D[\bar\phi,\phi]  \, e^{i S_{b} } e^{i \int \left ( J_{\bar \phi} \bar \phi +J_{\phi} \phi \right )}
                                                                     \, z_{[\bar \phi,\phi]} [ J_{\bar \psi} ,  J_{\psi} ] ,\\
  z_{[\bar \phi,\phi]} [ J_{\bar \psi} ,  J_{\psi} ] &= \int \text D[\bar \psi , \psi] \, e^{iS_{f}[\bar \psi, \psi] } \,e^{iS_{\rm int}[\bar \psi,\psi,\bar \phi,\phi]}
                                                       \, e^{i \int \left ( J_{\bar \psi} \bar \psi +J_{\psi} \psi \right )}.
\end{align}
\end{widetext}
Here we will focus on fermionic dynamical properties, so it is sufficient to look at $ \mc Z[  J_{\bar \psi} ,  J_{\psi} ] \equiv  \mc Z[ 0,0,  J_{\bar \psi} ,  J_{\psi} ] $.
The free boson equations of motion are $ i \partial_t \phi - \Omega  \phi = 0$ and $i \partial_t \bar \phi + \Omega  \bar \phi = 0 $.
There exists only one static solution to these free equations of motion \footnote{This is the analytic continuation to real times
  of the only solution to the imaginary-time equations of motion that satisfies the Matsubara periodicity conditions.}: $\phi(t) = 0 = \bar \phi(t)$.
Meanwhile, there are time-dependent solutions to these free equations of motion,
namely $\phi_{\rm sp}(t)=\phi_0 \,e^{-i\Omega t}$ and $\bar \phi_{\rm sp} (t)=\bar \phi_0 \,e^{+i\Omega t}$.
While \emph{statistical} properties of the system (given by the partition function, which is the analytic continuation to imaginary times of $\mc Z[ 0,0,0,0 ]$) are dominated by the static solutions,
\emph{dynamical} properties are also dominated by these time-dependent saddle-point solutions.
Here we use ``dominated'' in the usual sense that, when written explicitly in the basis of Fourier modes of the boson field,
\begin{align}
  \label{eq:2}
  \mc Z[ J_{\bar \psi} ,  J_{\psi} ]  &= \int \prod_{\omega \in \mathbb R}\frac{\text d \bar \phi_\omega \text d \phi_\omega }{2\pi i}   \, z_{[\bar \phi,\phi]} [ J_{\bar \psi} ,  J_{\psi} ] \nonumber \\
  & \qquad \times \exp \left \{i  \sum_{\omega \in \mathbb R}  \bar \phi_{\omega-\Omega} (\omega - \Omega) \phi_{\omega-\Omega} \right \} 
\end{align}
gets non-oscillatory (and thus dominant) contributions from the $\omega-\Omega=0$ zero-mode, i.e.\ the saddle-point solution.
We now perform a saddle-point approximation, which amounts to averaging the effect of transverse (i.e. $\omega \neq \Omega$) bosonic fluctuations
to obtain an effective generating functional for the fermions where only fluctuations within the saddle-point manifold are retained.
That is to say, effective fermionic dynamics will be computed as
\begin{align}
  \label{eq:168}
   \mc Z[ J_{\bar \psi} ,  J_{\psi} ]  &\approx \int \frac{\text d \bar \phi_0 \text d \phi_0 }{2\pi i} \, z^{\rm eff}_{(\bar\phi_0,\phi_0)} [ J_{\bar \psi} ,  J_{\psi} ] 
\end{align}
with the saddle-point reduced generator
\begin{align}
  \label{eq:43}
 z^{\rm eff}_{(\bar\phi_0,\phi_0)} [ J_{\bar \psi} ,  J_{\psi} ] &= 
                                                                                   \int \text D[\bar \psi , \psi] \, e^{iS_{f}[\bar \psi, \psi] } \, e^{i \int \left ( J_{\bar \psi} \bar \psi +J_{\psi} \psi \right )} \nonumber \\
  &\quad \times e^{iS_{\rm int}[\bar \psi(t),\psi(t), \bar \phi_0 \,e^{+i\Omega t}, \phi_0 \,e^{-i\Omega t}]} .
\end{align}
For the particular case of $S_{\rm int}= -\int \text dt\,H_{\rm int}$ given by Eq.\eqref{eq:30},
the latter phase factor identifies as $e^{-i\int \text d t V_{\rm eff}}$ with 
\begin{align}
  \label{eq:63}
  V_{\rm eff} =  \frac 1 2\sum_{\bs K \in {\rm HS}} \int'_{\bs k} \bar \psi_{\bs K}(\bs k) \,\sigma^x \Delta(\bs k,t)\, \psi_{\bs K}(\bs k) 
\end{align}
with $\Delta(\bs k,t)$ given by Eq.\eqref{eq:107}.

Even at this final stage, the bosonic field is not assumed to condense, and indeed averages out to zero,
since in Eq.\eqref{eq:43} one still performs the average over zero-mode amplitudes $(\bar \phi_0, \phi_0)$.
The latter averaging will be performed eventually to access $\mc Z[ J_{\bar \psi} ,  J_{\psi} ] $, meanwhile in a given $(\bar \phi_0, \phi_0)$ sector 
one only needs to study the effective physics of Eq.\eqref{eq:93}.

\section{Derivation of the effective tunneling geometry and dynamics}
\label{sec:geometry-dynamics-at}

\subsection{Geometry of the FS and crossing points}
\label{sec:pedestrian}

The geometry of the problem Eq.\eqref{eq:43} is the same as that of the Fermi surface reconstruction problem,
where one considers two intersecting Fermi surfaces, one of which is obtained by folding the other due to unit cell doubling, so that their centers are separated by $2\rm q = \sqrt2\pi/a$.
For simplicity we consider initially a circular FS with parabolic dispersion and mass $m_{\text \Circle}$.
Then the two ``band'' energies are
\begin{align}
  \label{eq:6}
  \epsilon_\pm(\bs k) = \frac 1 {2 m_{\text \Circle}} \big ( (k_x \mp {\rm q})^2 + k_y^2 \big ) ,
\end{align}
where locally within this appendix we use the axes' orientation as in Fig.\ref{fig:gap_geometry} (left).

The Fermi surfaces cross at coordinates $k_x=0$ and $k_y$ given by
\begin{align}
  \label{eq:7}
  \epsilon_\pm\big |_{k_x=0} = E_{\rm F} \; \Leftrightarrow \; k_y = \pm' \sqrt{k_{\rm F}^2 - {\rm q}^2}
\end{align}
where $k_{\rm F}=2 m_{\text \Circle} E_{\rm F}$.
Let us also notice that
\begin{subequations}
\begin{align}
  \label{eq:8}
  \tfrac 1 2 (\epsilon_+ - \epsilon_-) &= - k_x {\rm q}/m_{\text \Circle} , \\
   \tfrac 1 2 (\epsilon_+ + \epsilon_-) &=  (\bs k^2 + {\rm q}^2)/2 m_{\text \Circle} .
\end{align}
\end{subequations}

Now we include the coupling between the two FS such as worked out in Eq.\eqref{eq:75},
which going back to notations in the full Brillouin zone means we are considering the following two-by-two Hamiltonian matrix
\begin{align}
  \label{eq:9}
  \hat H_{\bs k} =  \frac{\bs k^2 + {\rm q}^2}{2 m_{\text \Circle} }\,\idmatrix - \frac{ k_x {\rm q}}{m_{\text \Circle}} \, \sigma^z + \Delta(\bs k) \sigma^x ,
\end{align}
where it is implicit that we consider a snapshot at a given fixed time $t$ and for fixed values of $\bar \phi_0,\phi_0$,
and that the effect of $\Delta(\bs k)$ will be negligible in any case away from the hot spots (i.e. crossing points),
as can be read off from the eigenvalues of $ \hat H_{\bs k} $ which are
\begin{align}
  \label{eq:10}
  E_\pm(\bs k)  =   \frac{\bs k^2 + {\rm q}^2}{2 m_{\text \Circle} } \pm \sqrt{(k_x{\rm q}/m_{\text \Circle})^2 + \Delta(\bs k)^2} .
\end{align}

Now we endeavour to determine the geometric parameters of the gap, Eqs.\eqref{eqs:1089}, depicted in Fig.\ref{fig:gap_geometry}.
For this, one needs to solve
\begin{align}
  \label{eq:11}
   E_\pm \big |_{k_x=0} = E_{\rm F} \; \Leftrightarrow \; k_y = \pm' \sqrt{2 m_{\text \Circle} (E_{\rm F}\mp\Delta) -{\rm q}^2} ,
\end{align}
which allows (by taking the difference between two $\pm$ solutions at fixed $\pm'$, then expanding to leading order in $\Delta$) to obtain $\delta k_y$, Eq.\eqref{eq:1089b}.
Now we can use the fact that locally near the crossing point, the two circular Fermi surfaces asymptotically become straight lines whose slope is both
$\delta k_x/ \delta k_y $ (see Fig.\ref{fig:gap_geometry}) and  $\sqrt{(k_{\rm F}/{\rm q})^2 - 1}$.
This determines $\delta k_x$, Eq.\eqref{eq:1089a}.
One can note that the FS geometry near the crossing point is described by a conic equation,
\begin{align}
  \label{eq:96}
k_x^2/\delta k_x^2 - k_y^2/\delta k_y^2 = -1 .
\end{align}
This will be useful below, App.\ref{sec:mapping-weber-eqns}.

\subsection{Coordinate mappings in a magnetic field}
\label{sec:mapping-weber-eqns}

We now consider a neighborhood of one crossing point, and redefine coordinates $\bs k$ to originate at the crossing.
Instead of the two-band hamiltonian $H_{\rm eff}$, we will start by considering a single-band saddle dispersion,
\begin{align}
  \label{eq:76}
   H_{\rm eff}^{\rm sb}(\bs k) = k_x^2/2m_x - k_y^2/2m_y ,
\end{align}
with yet unspecified parameters $m_x,m_y$.
Notice that by choosing the ratio $m_x/m_y = (\delta k_x/\delta k_y)^2$, the iso-energy curve $ H_{\rm eff}^{\rm sb}=\mc E$
at the energy $\mc E = \tfrac 1 2 \delta k_x \delta k_y / (m_xm_y)^{1/2}$ has precisely the form of Eq.\eqref{eq:96}: we will use this correspondence below.

For now we come back to Eq.\eqref{eq:76} and switch on a magnetic field, which is done in the Landau gauge by
transforming $k_x \rightarrow k_x - y/l_B^2$ in the Hamiltonian, thereby transforming $H_{\rm eff}^{\rm sb} \rightarrow \bar H_{\rm eff}^{\rm sb}$.
The Schrödinger equation, $\bar H_{\rm eff}^{\rm sb} \psi = \mc E \psi$,
upon redefinition $\tilde \psi = e^{-ik_xk_y}\psi$ as mentioned in the main text, and recalling $y = -i\partial_{k_y}$, reads
\begin{align}
  \label{eq:79}
  \left ( - \frac{l_B^{-4}}{2 m_x}\partial_{k_y}^2 - \frac 1 {2m_y} k_y^2 - \mc E \right ) \tilde \psi =0 .
\end{align}
This equation is that of a one-dimensional tunneling problem, which may be addressed in two different ways,
one which we explain in the following App.\ref{sec:acti-integr-form} and in the main text, and the other one which we discuss here.
One then performs the change of variables 
\begin{align}
  \label{eq:80}
   k_y = \frac{e^{i\pi/4}}{\sqrt 2 l_B}\left [ \frac{m_y}{m_x}\right ]^{\frac 1 4} \! r = \frac{e^{i\pi/4}}{\sqrt 2 l_B }  \sqrt{ \delta k_y/ \delta k_x}\, r ,
\end{align}
which together with $y = -i\partial_{k_y}$ playing the same role as $l_B^2k_x$ is equivalent to Eqs.\eqref{eq:99} in the main text.
This recasts Eq.\eqref{eq:79} into the form of Weber's differential equation,
\begin{align}
  \label{eq:81}
  \left ( \partial^2_{rr} - \frac 1 4 r^2 + i \mu \right ) \tilde \psi = 0 ,
\end{align}
with $\mu = \sqrt{m_xm_y}\,l_B^2\mc E = \frac 1 2 l_B^2 \delta k_x\delta k_y $. 
The asymptotic form of the solutions of Eq.\eqref{eq:81} is known, and allows to determine the scattering matrix
associated with this tunneling problem.\cite{aris-prb,aris-prl}

Now, how does the single-band $H_{\rm eff}^{\rm sb}$ in Eq.\eqref{eq:76} relate to the two-band $H_{\rm eff}$ at the crossing point?
Performing on $H_{\rm eff}$ the same analysis as above, one finds\cite{aris-prb,aris-prl} that similarly
the problem amounts to two differential equations for two functions $\tilde \psi_{l},\,l=1,2$, which are of the Weber form:
\begin{align}
  \label{eq:44}
  \left ( \partial^2_{rr} - \frac{r^2}{4} - (-1)^l\tfrac 1 2 + i \mu \right ) \tilde \psi_l = 0 .
\end{align}
The $(-1)^l\tfrac 1 2 $ extra term, different in both equations, is a specific feature of the two-band problem.
The asymptotic form of the solutions is again known, and determines the scattering matrix which is Eq.\eqref{eq:116}.
Crucially, while this is \emph{not} the same $\mathbb S(\mu)$ as in the single-band problem, the parameter $\mu$ itself is rigorously the same.
In this paper, we are concerned with determining the tunneling parameter $\zeta$ in a generalized setting where the ``band crossing'' is now governed by a fluctuating bosonic field,
but the algebraic structure of the tunneling problem (and thus the form of $\mathbb S(\zeta)$) is unchanged.
Therefore, we will be using $\mathbb S(\zeta)$ from Eq.\eqref{eq:116} together with a value of $\zeta$ determined by a different method, more adapted to our problem at hand.

It will be useful, for the action integral method described in App.\ref{sec:acti-integr-form}, to define the new variable $z = e^{i\pi/4}(4 m \kappa)^{-1/4} r$
and parameter $E = - \tfrac 1 2 (\kappa/m)^{1/2}\mu$ (note, this has nothing to do with $\mc E$ above),
where $m,\kappa$ are unspecified parameters whose role is purely pedagogical, as explained also in the main text.
With these replacements, the single-band equation Eq.\eqref{eq:81} is recast into
\begin{align}
  \label{eq:45}
  \left [ - \tfrac 1 {2m} \partial^2_{zz} - \frac{\kappa}{2} z^2 -E \right ] \tilde \psi = 0 ,
\end{align}
which is the equation of motion for tunneling of a spinless particle with mass $m$
through an inverted parabola $V(z)=- \frac 1 2 {\kappa}z^2$, at fixed energy $E<0$.
We stress that Eq.\eqref{eq:45} is a collection of problems, one for each value of $E$.
Our initial problem, of magnetic breakdown of electrons at an avoided crossing with geometric parameters $\delta k_x,\delta k_y$,
corresponds to just one problem in the collection, namely that at energy $E = - \tfrac 1 4 (\kappa/m)^{1/2} l_B^2 \delta k_x\delta k_y $.
We will check this explicitly in the next appendix.

\section{Action integral formulation of the effective tunneling problem}
\label{sec:acti-integr-form}

\subsection{Formalism for a general tunneling problem}
\label{sec:form-gener-tunn}

\begin{figure}[htbp]
  \begin{center}
    \includegraphics[width=.8 \columnwidth]{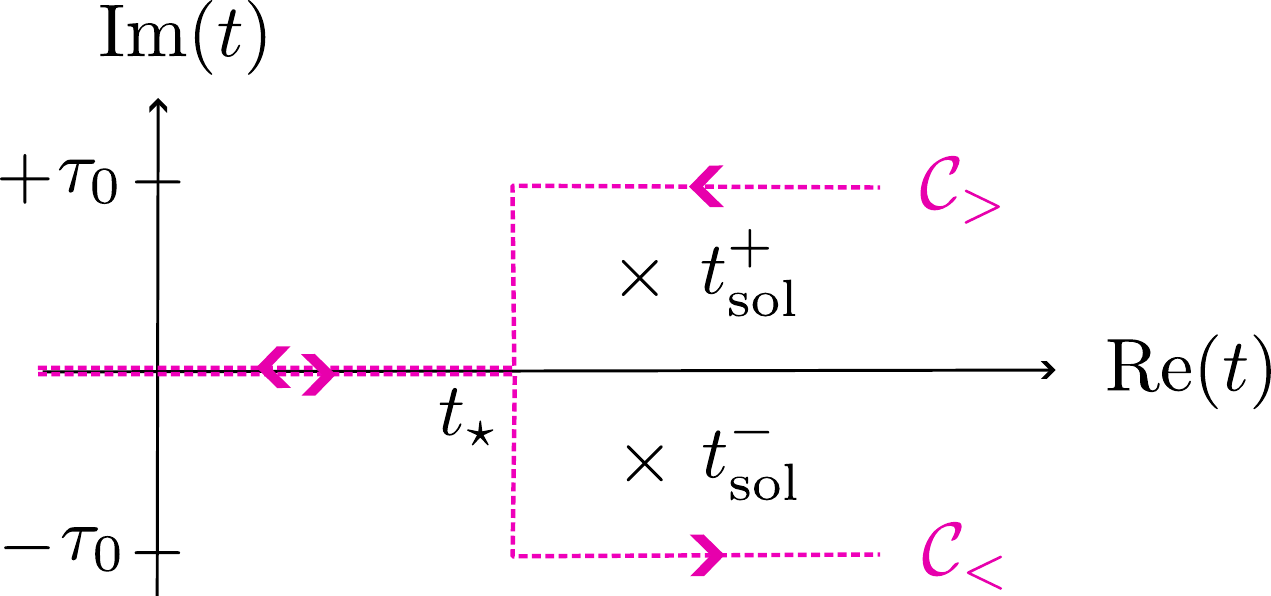}
    \caption{Complex time evolution along the contours $\mc C_<$ (forward) and $\mc C_>$ (backward).}
\label{fig:time-path}
\end{center}
\end{figure}

This appendix reproduces well-known results, see e.g. \cite{LL3,berry1972semiclassical,kamenev2011field}.
The classical Hamilton-Jacobi action for a particle evolving in two-dimensional phase space $(z,p)$, in an arbitrary potential $V(z)$,
following hamiltonian evolution with fixed energy $E$, is
\begin{align}
  \label{eq:46}
  \mc S(E) = \int_{\mc C_< + \mc C_>}\text d t \left ( p \dot z - \tfrac 1  2 m \dot z^2 - V(z) + E \right ).
\end{align}
The forward and backward time contours $\mc C_<,\mc C_>$ are depicted in Fig.\ref{fig:time-path} and described in detail below --
they are defined as those contours in the complex plane $t \in \mathbb C$ along which the Hamilton-Jacobi equation is satisfied (for a given boundary condition).

We consider tunneling through a barrier symmetric around position $z=0$, with classical turning points $z_\pm \equiv \pm z_\star$ defined by $V(z_\pm)=E$.
Let us choose an arbitrary origin at position $z_0< z_- $ located well before the barrier, so that at the level of the classical equations of motion,
\begin{subequations}
  \begin{align}
  \label{eq:47a}
    p & = m \frac{\text dz}{\text dt} = \sqrt{2m(E-V(z))} \\
    \label{eq:47b}
    \Leftrightarrow \qquad   t(z) &= \int_{z_0}^z \text d z'/\sqrt{2(E-V(z'))/m} ,
\end{align}
\end{subequations}
which defines $t(z)$ as the (complex) time it takes to reach position $z$ starting from $z_0$, following the classical equations of motion.
In particular let us note $t_\star = t(z_\star)$. The forward time contour $\mc C_<$, as depicted in Fig.\ref{fig:time-path}, can be determined as follows.
For $z \leq z_-$, time is $t \leq t_\star \in \mathbb R$.
Between $z_- \leq z \leq z_+$, time flows along the imaginary direction, from $t_\star$ to $t_\star-i\tau_0$ defined by
\begin{align}
  \label{eq:48}
  \tau_0 = \int_{z_-}^{z_+}\text d z'/\sqrt{2(V(z')-E)/m} .
\end{align}
For $z \geq z_+$, time is $t -i \tau_0$,  $t \geq t_\star \in \mathbb R$. 
The backward time contour $\mc C_>$ can be determined in a similar fashion: see Fig.\ref{fig:time-path}.
Computing the classical action (henceforth denoted $\mc S_0$), only the integral along the imaginary axis does not vanish,
\begin{align}
  \label{eq:49}
  \mc S_0 = \int_{t_\star + i\tau_0}^{t_\star - i\tau_0} \text dt \, m\dot z^2 = 2i \int_{z_-}^{z_+}\text d z \, \sqrt{2m(V(z)-E)}.
\end{align}
The transmission coefficient is then $e^{-\pi \mu} \;\equiv\;  e^{i \mc S_0}$, which defines $\mu$ for any (static) tunneling problem at a given energy $E$.

\subsection{Specifically for an inverted parabola}
\label{sec:spec-an-invert}

In the specific case of an inverted parabolic potential $V(z)=- \frac{1}{2} \kappa z^2$,
at negative energies $E<0$ tunneling happens between points $z_\pm=\pm z_\star $ where $z_\star = \sqrt{-2E/\kappa}$.
Then one can compute explicitly, using Eq.\eqref{eq:49}, $\mu = z_\star^2 \sqrt{\kappa m}$.
In other words, this relates the energy $E$ to the static tunneling parameter $\mu$ as
\begin{align}
  \label{eq:52}
  E = - \tfrac 1 2 \kappa z_\star^2 = - \tfrac 1 2 (\kappa/m)^{1/2} \mu,
\end{align}
in full consistency with the relation given about Eq.\eqref{eq:45}. This means that the tunneling parameter $\mu$, as obtained in the action integral formalism,
is indeed the same as that appearing in the different language of Weber's differential equation, Eq.\eqref{eq:45}.
We note that it is also straightforward to compute explicitly, using Eq.\eqref{eq:48}, the tunneling time $ \tau_0 = \pi (m/\kappa)^{1/2}$,
whence the relation $\mc S_0 = -2i E \tau_0$ which illustrates very clearly that only the tunneling event contributes to the action.

We finally note, for later use, the solution to the classical equations of motion $\ddot z = (\kappa/m) z $ along the forward contour $\mc C_<$:
\begin{subequations}
  \label{eq:55}
\begin{align}
  \label{eq:55a}
  z(t) &= z_0 \cosh( \sqrt{\kappa/m} (t-t_0)) \\
  \label{eq:55b}
  &= - z_\star \cosh( \sqrt{\kappa/m} (t-t_\star)) .
\end{align}
\end{subequations}
Beside $ z(t_\star)=-z_\star $ one can also check $ z(t_\star - i\tau_0) = - z_\star \cosh(-i\pi) = +z_\star $, as it should.

\subsection{Case of a time-dependent Lagrangian}
\label{sec:time-depend-oscill}

We now consider the case described in Eq.\eqref{eq:110} where the potential $V(z)$ is generalized to a time-dependent Lagrangian $V(z,p,t)=V_0(z)+\varepsilon V_1(z,p,t)$,
with deviations from the static potential case treated as a small perturbation.
The case with a $t$-dependence but without a $p$-dependence was investigated in Ref.\cite{ivlev-melnikov}: we generalize their analysis here.

There will be two modifications in the action Eq.\eqref{eq:46} arising from the new contribution $\varepsilon V_1(z,p,t)$ to the particle's Lagrangian:
\begin{itemize}
\item[-] The time contours $\mc C_< ,\mc C_>$, in particular the turning points $z_\pm$, and the definition of the conjugate momentum $p=\partial \mc L/\partial \dot z$ will change. 
\item[-] There is a completely new term in the action,
  \begin{align}
    \label{eq:76}
    \mc S_1(E)= - \varepsilon \int_{\mc C_< + \mc C_>}\text d t \, V_1(z,p,t) ,
  \end{align}
  where to the order $\varepsilon^1$ it is sufficient to keep the static problem's time contours $\mc C_< ,\mc C_>$ and phase space coordinates $z(t),p(t)$.
\end{itemize}
The former effect is usually neglected \cite{kamenev2011field}, which we will also do here, focussing on the latter contribution.
We thus want to compute
\begin{align}
  \label{eq:179}
 \varepsilon\mc S_1 &= -  \varepsilon E_0 \int_{\mc C_< + \mc C_>}\text d t \, \,  \frac{ \cos\left ( 2\Omega t - 2 \alpha_0 \right )} {1+ (z/\xi)^2 - (p /m v)^2} 
\end{align}
where $z(t)$ is given by Eqs.\eqref{eq:55} and simply $p(t)=m\dot z(t)$.
Below, we use the shorthand
\begin{align}
  \label{eq:80}
  K^{-2} &=\frac{ (z_\star/\xi)^2 -  (z_\star/v  \tau_0 )^2 }{1  +   (z_\star/v  \tau_0)^2} ,
\end{align}
assuming $K^{-2}>0$ i.e.
\begin{align}
  \label{eq:83}
  1 < v  \tau_0 /\xi = \pi (\Pi_x/\Pi_y)(k_{\rm F}^2/{\rm q}^2-1)^{-1/2} ,
\end{align}
reasonable since ${\rm q} < k_{\rm F}<\sqrt 2 {\rm q}$, provided $\Pi_x \approx \Pi_y$.
\begin{widetext}
Absorbing the phase $2\alpha_0$ into a redefinition of the origin $t=0$, we just need to compute
\begin{align}
  \label{eq:38}
 \mc S_1 = - \frac{E_0}{1 +   (z_\star/v  \tau_0)^2}  \int_{\mc C_< + \mc C_>}\text d t \,
  \frac{\cos(2\Omega t)}{ 1 + K^{-2} \cosh^2(\sqrt{\kappa/m}(t-t_\star))  } .
\end{align}
Thus, the denominator vanishes whenever
\begin{align}
  \label{eq:58}
  \cosh(\sqrt{\kappa/m}(t-t_\star)) = \pm i K \quad \Leftrightarrow \quad t = t_\star \pm \sqrt{m/\kappa} \big ( \text{arcsinh}(K) + i \pi ({\rm n}+1/2) \big ),\;{\rm n}\in \mathbb Z .
\end{align}

\end{widetext}
The integral Eq.\eqref{eq:38} can be computed using the residue theorem. The only poles contained within the time contour are
\begin{align}
  \label{eq:5}
  t_{\rm sol}^\pm = t_\star + \sqrt{m/\kappa} \big ( \text{arcsinh}(K) \pm i \pi/2  \big ),
\end{align}
with residue $\tfrac 1 2 \big (1+K^{-2}\big )^{-1/2}$ for both.
The upshot is
\begin{align}
  \label{eq:59}
  i \mc S_1 = +\frac{E_0}{1 +   (z_\star/v  \tau_0)^2}  \frac{2 \tau_0 \,\cos(\lambda)\, \cosh(\Omega \tau_0)}{\sqrt{1+ K^{-2}}}
\end{align}
with $\lambda = 2\Omega t_\star - (2\Omega \tau_0/\pi) \, \text{arcsinh}(K) $.
This is the result quoted in the main text, Eq.\eqref{eq:111}.
We note that in the limit $v  \rightarrow \infty$, where $p$-dependence is removed from $\varepsilon V_1(z,p,t)$, this recovers the known result (see \cite{kamenev2011field}).
For practical use, we report the explicit expressions:
\begin{align}
  \label{eq:83}
  z_\star/v\tau_0 &= \tfrac 1 \pi v_{\rm F}^{-1}\sqrt{\sigma} \Delta_0 \Pi_x^{-1}(k_{\rm F}^2/{\rm q}^2-1)^{1/4} ,\\
 z_\star/\xi &= v_{\rm F}^{-1}\sqrt{\sigma} \Delta_0 \Pi_y^{-1}(k_{\rm F}^2/{\rm q}^2-1)^{-1/4} .
\end{align}

\section{Physical observables from averaging over the bosonic field}
\label{sec:phys-observ-aver}

\subsection{Deriving the averaged tunneling probability}
\label{sec:deriv-aver-tunn}

First we provide arguments for Eq.\eqref{eq:117}.
Naively, the average over the complex bosonic amplitude $\phi_0=\sqrt{\nu_0}e^{i\alpha_0}$is not weighed by temperature, for instance
$${\mc Z } = \int \frac{\text d \phi_0^* \text d \phi_0}{2\pi i} = \int_0^\infty \text d \nu_0 \int_0^{2\pi}\frac{\text d \alpha_0}{2\pi} =  \int_0^\infty \text d \nu_0$$
produces a divergence, as is usual in zero-temperature functional integrals.
This can be traced back to our not enforcing the Matsubara periodicity relations as we performed a real-time saddle-point approximation, see App.\ref{sec:saddle-point-appr-1}.
A physically meaningful way to re-introduce temperature is to quantize the bosonic zero-mode,
turning the integral into a sum and restoring Boltzmann weights, so that for any integrand function $I$ one replaces
\begin{align}
  \label{eq:182}
  \int_0^\infty \text d \nu_0\, I(\nu_0)\quad \rightarrow \quad \sum_{\nu_0=0}^\infty e^{-\nu_0 \Omega / T}\, I(\nu_0) .
\end{align}
Notice that taking $I = 1$ in the above implies that $\mc Z=\left ( 1 - e^{-\Omega/T} \right )^{-1}$ is the correct bosonic partition function.

Now we provide details of the derivation of Eq.\eqref{eq:102}.
Keeping only the first-order term in $\zeta=\zeta_0+\varepsilon \zeta_1+\dots$, and restoring $\varepsilon \rightarrow 1$, one wants to evaluate
\begin{align}
  \label{eq:181}
  \langle\rho^2\rangle &\equiv \langle e^{-2\pi \zeta} \rangle = \frac 1 {\mc Z }\int \frac{\text d\phi_0^*\text d \phi_0}{2\pi i}\, e^{-2\pi ( \zeta_0+\zeta_1)} .
\end{align}
First we perform the phase average over $\alpha_0$, since (see Eq.\eqref{eq:111}) one has $\zeta_1 \propto \cos(\lambda)$
and $\lambda = 2\alpha_0 + \dots$ (recall the origin $t=0$ had absorbed an offset $\alpha_0/\Omega$). 
The integral amounts to $\int_0^{2\pi}\frac{\text d \alpha}{2\pi} \exp \left [\beta \cos(2\alpha) \right ] = {\sf I}_0(\beta)$ where ${\sf I}_0$ is the zeroth modified Bessel's function.
Then one sums over the mode amplitude. Using the physical prescription Eq.\eqref{eq:182} above, equivalently Eq.\eqref{eq:117},
eventually Eq.\eqref{eq:181} becomes Eq.\eqref{eq:102} in the main text.

\subsection{Technical comment on convergence}
\label{sec:techn-comm-conv}

Because Eq.\eqref{eq:102} was obtained from a perturbative expansion of the action, $\mc S = \mc S_0 + \varepsilon \mc S_1 + \dots$,
one has to make sure that the truncation to the order $\varepsilon^1$ is reasonable -- in particular, that $\zeta_1$ does not get larger than $\zeta_0$.
It is known \cite{ivlev-melnikov,kamenev2011field}, and can be easily checked, that the limit of large $\Omega$ does not satisfy this requirement,
because the cosh factor in $\mc S_1$, Eq.\eqref{eq:59}, diverges while $\mc S_0$ remains finite.
In fact, this also entails that the sum in Eq.\eqref{eq:102}, for any $\Omega$, diverges: this is a serious technical issue that must and can be fixed.

This sum is of the form
\begin{align}
  \label{eq:84}
 S = \sum_{\nu=0}^{\infty} e^{-\nu \Omega/T} e^{- C\nu} {\sf I}_0 \big ( C \nu R(\nu) \big ),
\end{align}
where $R(\nu)$ is some function (namely the ratio of a cosh and a square root in Eq.\eqref{eq:102})
that clearly satisfies $R(0)=1$ and $R(\infty)=\infty$ with exponential growth.
Given the asymptotic behavior ${\sf I}_0(s)\sim e^s/\sqrt{2\pi s}$ at $s\rightarrow \infty$, such a sum is always strongly divergent.
However, one can see, for instance by expanding the numerator and denominator of $R(\nu)$ to the first order in $\nu$,
that there exists a critical value of $\Omega$ below which $R(\nu)$ first decreases, remaining smaller than 1 over a range of values of $\nu$, before eventually blowing up.
For such values of $\Omega$, the summand of $S$ first decays exponentially rapidly with $\nu$,
so that by truncating the sum to a finite (possibly large) number of terms, it appears to have converged;
only after summing a yet much larger number of terms does one witness the eventual divergence.
Then, the (physically relevant) regularization scheme simply consists of truncating the series at such values where it has apparently converged.
This is how we obtained the results displayed in Fig.\ref{fig:plots}.
We note that this procedure is not original, as it is foundational in quantum field theory \cite{dyson1952divergence,hurst1952example}.
It is also, importantly, \emph{physically} justified, as it amounts to removing spurious effects of the truncation to the order $\varepsilon^1$ in the region where it is insufficient
(see Ref.\cite{kamenev2011field} for a pedagogical discussion), keeping only those contributions in the region where it can be relied upon.

\subsection{Possibility of thermal damping induced by magnetic breakdown}
\label{sec:techn-deta-therm}

Here we justify the statement made at the end of Sec.\ref{sec:connection-formula},
that thermal fluctuations of $\omega,\rho$ do not, in general, entail any significant damping of the quantum oscillations.

First, we note that $\omega \in [-\pi/4,0]$ is bounded, and moreover the quantization condition Eq.\eqref{eq:1066}
simply does not depend on $\omega$ in the relevant limit of small $\zeta$.
Therefore, thermal fluctuations of $\omega$ may not generate any significant damping of quantum oscillations.

We thus focus on the effect of thermal fluctuations of $\rho$.
To quantify their effect on the semiclassical phase fluctuations, one may look at the derivatives
on the manifold of solutions $(\vartheta,\varphi)$ of Eq.\eqref{eq:1066},
namely $\text d \vartheta \big /\text d \rho, \text d \varphi\big /\text d \rho$, which by dint of the implicit function theorem are
\begin{align}
  \label{eq:32}
  \frac{\text d \vartheta}{\text d \rho} = - \frac{\partial \mc F \big / \partial \rho}{\partial \mc F \big / \partial \vartheta} \bigg |_{\mc F = 0} \;\text {and} \;
  \frac{\text d \varphi}{\text d \rho} = - \frac{\partial \mc F \big / \partial \rho}{\partial \mc F \big / \partial \varphi}\bigg |_{\mc F = 0}  .
\end{align}
It is easy to check numerically that these derivatives remain of order $O(1)$ for almost all values of $\zeta$ (which fixes $\rho,\omega$).
Since $\rho \in [0,1]$ is bounded, thermal fluctuations of $\rho$ may not generate any significant damping of quantum oscillations either.
\footnote{Strictly speaking, a strong enhancement of this thermal damping may occur for a discrete set of values of $\zeta$: we do not explore this possibility further here.}

\section{Derivation of the WKB connection formula}
\label{sec:conn-form-cupr}

\subsection{Summary of amplitude relations}
\label{sec:summ-all-ampl}

As an initial caveat, we note that the scattering formula Eq.\eqref{eq:115} is not left-right symmetric,
indeed ${\mathbb S} \neq  \hat \sigma^x {\mathbb S} \hat \sigma^x $.
Instead, we have
\begin{align}
  \label{eq:17}
  {\mathbb S}^\dagger = {\mathbb S}^{-1} = \hat \sigma^x {\mathbb S} \hat \sigma^x ,
\end{align}
so the formula is symmetric upon rotation of the picture in Fig.\ref{fig:gap_geometry}, i.e.\ exchanging both $L/R$ and $+/-$.
This formula implies relations between the wavefunction amplitudes depicted on Fig.\ref{fig:schema} at the four crossing points:
\begin{align}
  \label{eq:20}
 & \begin{pmatrix}  a_3 \\ \bar a_4 \end{pmatrix} =   {\mathbb S}  \begin{pmatrix}  a_4 \\ \bar a_3  \end{pmatrix} ,\quad
  \begin{pmatrix}  a_7 \\ \bar a_8  \end{pmatrix} =  {\mathbb S}  \begin{pmatrix}  a_8 \\ \bar a_7  \end{pmatrix} ,\nonumber \\
 & \begin{pmatrix}  \bar a_1 \\  a_2  \end{pmatrix} =  {\mathbb S}^{-1}  \begin{pmatrix}  \bar a_2 \\ a_1  \end{pmatrix} ,\quad
  \begin{pmatrix}  \bar a_5 \\  a_6  \end{pmatrix} =  {\mathbb S}^{-1}  \begin{pmatrix}  \bar a_6 \\ a_5  \end{pmatrix} .
\end{align}

Besides, the semiclassical evolution along the arc-shaped portions of path amounts trivially to phase factors $e^{i\vartheta}$ or $e^{i\varphi}$.
Grouped by pairs in a convenient way, these phase relations are
\begin{align}
  \label{eq:19}
&  \begin{pmatrix}  a_1 \\ \bar a_7  \end{pmatrix} =  \hat I_{\vartheta \varphi}  \begin{pmatrix}  a_3 \\ \bar a_4  \end{pmatrix} ,\quad
  \begin{pmatrix}  a_5 \\ \bar a_2  \end{pmatrix} =  \hat I_{\vartheta \varphi}  \begin{pmatrix}  a_7 \\ \bar a_8  \end{pmatrix} ,\nonumber \\
 & \begin{pmatrix}  \bar a_6 \\  a_4  \end{pmatrix} =  \hat I_{\varphi \vartheta}  \begin{pmatrix}  \bar a_1 \\ a_2  \end{pmatrix} ,\quad
  \begin{pmatrix}  \bar a_3 \\  a_8  \end{pmatrix} =  \hat I_{\varphi \vartheta}  \begin{pmatrix}  \bar a_5 \\ a_6  \end{pmatrix} ,
\end{align}
where one defined the useful notations
\begin{align}
  \label{eq:18}
  \hat I_{\vartheta\varphi} =
  \begin{pmatrix}
    e^{-i\vartheta}&0\\ 0 & e^{-i\varphi}
  \end{pmatrix}, \quad
                            \hat I_{\varphi \vartheta} =
  \begin{pmatrix}
    e^{-i\varphi}&0\\ 0 & e^{-i\vartheta}
  \end{pmatrix} .
\end{align}

\subsection{Obtaining the semiclassical quantization condition}
\label{sec:obta-semicl-quant}

Physical electronic ``modes'' of the system, as described above in Eqs.\eqref{eq:20},\eqref{eq:19}, may be understood intuitively as standing waves of a circuit,
in the more explicit sense that the collection of amplitudes $(a_i,\bar a_i), i=1..8$, must be an eigenvector of the unitary dynamics with eigenvalue 1.
It turns out that instead of a $16\times 16$ problem, a well-chosen subset that contains the full information about the system
allows one to deal with an $8\times 8$ problem instead:
\begin{equation}
  \label{eq:21}
  \begin{pmatrix}   a_1 \\ \bar a_7 \\ a_5 \\ \bar a_2  \\  \bar a_6 \\  a_4 \\  \bar a_3 \\  a_8 \end{pmatrix}
  = \underset{=\;\hat M}{\underbrace{
\begin{pNiceArray}[columns-width=4.5mm]{cccccccc}
  \Block{2-2}{ \hat I_{\vartheta \varphi}\mathbb S} && \Block{2-2}{ 0 } && \Block{2-2}{ 0 } && \Block{2-2}{ 0 } &\\
 &&&&&&& \\
\Block{2-2}{ 0 } && \Block{2-2}{ \hat I_{\vartheta \varphi} \mathbb S} &&\Block{2-2}{ 0 } && \Block{2-2}{ 0 } &\\
 &&&&&&&\\
\Block{2-2}{ 0 } && \Block{2-2}{ 0 } && \Block{2-2}{ \hat I_{ \varphi \vartheta}  \mathbb S^{-1}} &&\Block{2-2}{ 0 } &\\
 &&&&&&&\\
\Block{2-2}{ 0 } && \Block{2-2}{ 0 } && \Block{2-2}{ 0 } && \Block{2-2}{ \hat I_{\varphi \vartheta }  \mathbb S^{-1}} &\\
&&&&&&&
\end{pNiceArray}
}}
 \begin{pmatrix}   a_4 \\ \bar a_3 \\ a_8\\ \bar a_7 \\  \bar a_2 \\  a_1 \\  \bar a_6 \\  a_5 \end{pmatrix} ,
\end{equation}
where each element of the $4\times 4$ matrix is a $2\times 2$ block.
Rearranging the elements in the right-hand column into the same order as they are on the left-hand side, 
\begin{equation}
  \label{eq:22}
  \begin{pmatrix}   a_4 \\ \bar a_3 \\ a_8\\ \bar a_7 \\  \bar a_2 \\  a_1 \\  \bar a_6 \\  a_5 \end{pmatrix}
  =\underset{=\;\hat Q}{\underbrace{
    \begin{pmatrix}
      &&&&&1&&\\
      &&&&&&1&\\
      && &&&&&1\\
      &1&&&&&&\\
      &&&1&&&&\\
      1&&&&&&&\\
      &&&&1&&&\\
      &&1&&&&&
    \end{pmatrix}
               }}
     \begin{pmatrix}   a_1 \\ \bar a_7 \\ a_5 \\ \bar a_2  \\  \bar a_6 \\  a_4 \\  \bar a_3 \\  a_8 \end{pmatrix} ,
\end{equation}
one obtains the semiclassical quantization condition:
\begin{align}
  \label{eq:23}
  {\rm det}\left [ \hat M \hat Q - \idmatrix \right ] = 0 .
\end{align}
With the particular form of $\mathbb S$ given in Eq.\eqref{eq:116} in the main text, this becomes Eq.\eqref{eq:1066}.

Note that although we used the system's symmetries to express $\hat M \hat Q$ in terms of only three elementary ingredients
$\vartheta,\varphi$ and $\mathbb S$, our derivation applies equally well to any system with the same \emph{topology}
as that depicted in Fig.\ref{fig:schema}, with possibly different phases carried by all the ``links'' and different scattering matrices at all the ``junctions''.

\end{document}